\documentclass[tightenlines,superscriptaddress,eqsecnum,floats,nofootinbib]{revtex4}

\usepackage{amssymb,amsmath,amssymb,amsfonts,amsthm,stmaryrd,mathrsfs,physics,bbm}
\usepackage{graphicx}
\usepackage[T1]{fontenc}\usepackage{xcolor}
\usepackage{ulem}

\usepackage{hyperref} 
\newcommand{\makeSymbol}[1]{\mathord{\vcenter{\hbox{#1}}}}

\newtheorem{defi}{Definition}[section]

\newcommand{\Pl}{\ell_p}
\newcommand{\sut}{{\mathrm{SU(2)}}}
\newcommand{\cyl}{{\mathbf{Cyl}}}
\newcommand{\cylf}{{\mathbf{Cyl_F}}}
\newcommand{\cylt}{{\mathbf{Cyl_{tot}}}}
\newcommand{\grade}{{\mathfrak d}}
\newcommand{\diff}{{\rm Diff}}
\newcommand{\gs}{{\rm GS}}

\begin{document}

\title{ Fermion coupling to loop quantum gravity: canonical formulation. }

\author{Jerzy Lewandowski \footnote{Jerzy.Lewandowski(AT)fuw.edu.pl}}
\author{Cong Zhang\footnote{ czhang(AT)fuw.edu.cn}}
\affiliation{Faculty of Physics, University of Warsaw, Pasteura 5, 02-093 Warsaw, Poland}
\begin{abstract}
In the model of a fermion field coupled to loop quantum gravity, we consider the Gauss and the Hamiltonian constraints. According to the explicit solutions to the Gauss constraint, the fermion spins and the gravitational spin networks intertwine with each other so that the fermion spins contribute to the volume of the spin network vertices. For the Hamiltonian constraint, the regularization and quantization procedures are presented in detail. By introducing an adapted vertex Hilbert space to remove the regulator, we propose a diffeomorphism covariant graph-changing Hamiltonian constraint operator of the fermion field. This operator shows how fermions move in the loop quantum gravity spacetime and simultaneously influences the background quantum geometry.  Moreover, as an innovation of our work,  introducing the vertex Hilbert space also fixes issues so that a densely defined symmetric Hamiltonian constraint operator can be  obtained. 
\end{abstract}


\maketitle
\section{introduction}
The real physical world consists of spacetime and matter. According to general relativity (GR), \textit{``spacetime tells matter how to move; matter tells spacetime how to curve"-John Archibald Wheeler}, which should also be carried out in the quantum theory. Loop quantum gravity (LQG) \cite{ashtekar1991lectures,ashtekar2004back,rovelli2005quantum,thiemann2007modern,han2007fundamental}, as a background-independent and non-perturbative quantum gravity theory, sets a stage for incorporating matters into quantum spacetime. In \cite{rovelli1994physical,brown1995dust,giesel2010algebraic,domagala2010gravity,lewandowski2011dynamics}, the Brown-Kuchar model of gravity coupled to dust as well as the Rovelli-Smolin model of gravity coupled to massless Klein-Gordon field is quantized. In \cite{han2013spin,bianchi2013spinfoam}, a minimal coupling of fermions and Yang-Mills fields to covariant LQG dynamics is proposed. The quantum theory of spinor fields coupled to LQG is well understood \cite{morales1994fermions,morales1995loop,baez1998quantization,thiemann1998kinematical,bojowald2008fermions,barnett2015fermion,gambini2015no}.
In \cite{thiemann1998quantumV}, a systematic procedure to couple standard model to the canonical LQG is proposed and further developed in \cite{thiemann1998kinematical,mansuroglu2021fermion,mansuroglu2021kinematics}. With the present paper, by employing the procedure in \cite{thiemann1998quantumV}, we investigate the Gauss and the Hamiltonian constraint in the model of fermion field coupled to LQG. In particular, the Gauss constraint is solved explicitly and, the Hamiltonian constraint is regularized and quantized by introducing the so-called vertex Hilbert space to remove the regulator. 

In the classical model of gravity coupled to the fermion field. The gravity action $S_G$ can be formulated optionally with the first-order formulation (see, e.g., \cite{ashtekar1991lectures,holst1996barbero} for the Palatini-Host action) or the second-order formulation (see, e.g., \cite{arnowitt1959dynamical,thiemann2007modern} for Hilbert-Einstein action). In the pure gravity case, these two formulations are equivalent to each other up to boundary terms, while for the case with the fermion field coupled, the equivalency is no longer valid. For the first-order formulation, $S_G$ is a functional of an SL($2,\mathbb C$) connection, and the fermion field will be coupled to it directly. As a consequence, the fermion field will result in an on-shell torsion term in the connection. However, for the second-order formulation, the fermion field will be coupled to the torsion-free spin connection compatible with the tetrad. Thus, there is no torsion involved in this formulation. In the current paper, even though we adopt the second-order formulation for discussion, the results for the first-order formulation can be obtained analogously. Moreover, since no extra field is introduced for deparametrization in our model, the dynamics will be governed by the Hamiltonian constraint $H[N]$ with lapse functions $N$ rather than the physical Hamiltonian. Then a problem arises that the Hamiltonian constraint operator cannot be defined in the diffeomorphism invariant space. This problem will be solved by, for instance, the master constraint framework \cite{thiemann2006phoenix,han2006master} or the deparametrization framework \cite{domagala2010gravity,lewandowski2011dynamics}. In these frameworks, one finally needs to consider the Hamiltonian constraint operator with a constant lapse function or some dynamical lapse function. These operators can be constructed directly with the Hamiltonian constraint operators $\widehat{H[N]}$. Thus, one can only focus on how to define well the Hamiltonian constraint operators $\widehat{H[N]}$ without loss of generality,  which is a main task of the current work.

The phase space of the fermion field coupled to gravity is composed of fields $(A_a^i,E^b_j,\Psi,\Pi)$ on the spatial manifold $\Sigma$, where $A_a^i$ is an SU(2) connection, $E^b_j$, the canonical conjugate to $A_a^i$, is a densitized triad field, $\Psi$ denotes the fermion field, and $\Pi$ is the canonical momentum conjugate to $\Psi$ \cite{ashtekar1991lectures,barbero1995real,ashtekar1989new}. With the variables $(A_a^i, E^b_j)$, the gravitational Hilbert space is constructed by using the spin networks \cite{ashtekar1994representation,lewandowski1994topological,ASHTEKAR1995191}.
Therein,
the Hamiltonian constraint of pure gravity as well as  variables representing the geometric observables are regularized and promoted to operators, and properties of the operators are welll-studied \cite{rovelli1995discreteness,ashtekar1997quantum,ashtekar1997quantumII,thiemann1998quantum,thiemann1998quantumII, ma2000qhat,yang2016new,thiemann2006phoenix,han2006master,lewandowski2014symmetric,yang2015new,zhang2020first,zhang2021firstII}. The pure-gravity Hamiltonian constraint comprises of the curvature of the connection $A$. The curvature, as an operator, will attach loops on graphs of the spin network states. In some regularization strategies, the final operator is the limit as the loops shrink to a point. Then, the vertex Hilbert space is  necessarily introduced to define the limit \cite{alesci2010regularization,alesci2015hamiltonian,assanioussi2015new,zhang2018towards,zhang2019bouncing}.
 In general, a vertex Hilbert space is a Hilbert space averaged with diffeomorphisms preserving some particular vertices. Thus, elements in a vertex Hilbert space are partially diffeomorphism invariant.  In the current work, the Hamiltonian constraint of fermion is also regularized by introducing some regulator so that the fermion Hamiltonian constraint is the limit of the regularized  version as the regulator approaches $0$. In order to define this limit, we also need  to introduce an adapted vertex Hilbert space. Besides, the vertex Hilbert space can do more than just defining limit in this work. As shown in Sec. \ref{sec:Hoperators}, the regularized fermion Hamiltonian constraint is split into  $H_{\mathcal C_\epsilon}$ plussing its complex conjugate. Then, $H_{\mathcal C_\epsilon}$ can be quantized into an operator $\widehat{H(\delta)}$ which is cylindrical consistent and diffeomorphism covariant. However, the adjoint of $\widehat{H(\delta)}$, denoted by $\widehat{H(\delta)}{}^\dagger$, is not densely defined. Actually, as shown in Sec. \ref{sec:Hoperators}, acting on some particular states associated to a graph $\gamma$, the operator $\widehat{H(\delta)}$ can change $\gamma$ by erasing a segment $e(\delta)$ of an edge $e\subset\gamma$. Using $\gamma'$ to denote the graph of $\gamma$  with $e(\delta)$ erased by $\widehat{H(\delta)}$, $\widehat{H(\delta)}{}^\dagger$, acting on a state $\Psi_{\gamma'}$ associated to $\gamma'$, will add an edge $\tilde e(\delta)$ to $\gamma'$, where $\tilde e(\delta)$ is arbitrary  edge connecting the end points of $e(\delta)$. Thus, there will be infinitely many ways to add  $\tilde e(\delta)$ to $\gamma'$, and the resulting  states are orthogonal to each other. Consequently, the result of $\widehat{H(\delta)}{}^\dagger\Psi_{\gamma'}$ is not normalizable, i.e., $\Psi_{\gamma'}$ is not in the  domain of $\widehat{H(\delta)}{}^\dagger$. This problem will be fixed by introducing the vertex Hilbert space to define limit. In the vertex Hilbert space, $\lim_{\delta\to 0}\widehat{H(\delta)}$  will be  defined properly as an operator $\hat A_F^\dagger$. The action of $\hat A_F^\dagger$ performs in a way that  a projection is left multiplied in $\hat A_F^\dagger$ as a factor. Then, if $\hat A_F^\dagger$, acting on these particular states, erases segments, the projection will annihilate the resulting states, so that the adjoint of $\hat A_F^\dagger$ is densely defined. Actually, the same issue also exists in defining the gravitational Hamiltonian constraint operator with certain regularization strategy,   and can be fixed with the same mechanism \cite{yang2015new}.

The usual Hamiltonian analysis tells $\Pi^\dagger=\sqrt{q}\,\Psi$ with $q\equiv |\det(E)|$. In the quantum theory, this equation is expected to be realized in an appropriate form. In other words, one might require that the adjoint of the operator $\hat \Pi$ is related to the operator $\hat\Psi$ via $\hat\Pi^\dagger=\widehat{\sqrt{q}}\hat\Psi$. Then, contradiction  appears because in our quantum theory $\sqrt{q}$ will become an operator rather than some background c-number. This can be seen as follows. Considering a non-trivial real-valued function $f(A)$ of the connection $A_a^i$, one has $0=[\hat \Pi,\widehat{f(A)}]$ on one hand, but, on the other hand, $[\hat \Pi,\widehat{f(A)}]^\dagger=[\widehat{f(A)},\hat\Pi^\dagger]=[\widehat{f(A)},\widehat{\sqrt{q}}\hat\Psi]\neq0$. The two results do not coincide. To overcome this inconsistency, the author in \cite{thiemann1998quantumV} introduces the Grassman-valued half-densities $\tilde\Psi:=\sqrt[4]{q}\Psi$ and $\tilde\Psi^\dagger$ to define the phase space of fermion. Moreover, in order to do quantization, the smeared version of $\tilde\Psi$ by $\sqrt{\delta(x,y)}$, i.e., $\int\dd^3 y\sqrt{\delta(x,y)}\tilde\Psi(y)$, is also introduced by \cite{thiemann1998quantumV}. With the smeared variables, the fermion sector is quantized and the resulting quantum theory carries out the dieffeomorphsm-invariance feature.

It will be useful to compare our work with the known models of fermions in LQG. Among these models, a typical one is to introduce the path observables (see, e.g., \cite{morales1994fermions,morales1995loop,baez1998quantization}) each of which combines the holonomy along a path  $\alpha$ and the fermions at the ends of $\alpha$. These path variables form a closed algebra under the Poisson bracket so that the quantum theory is obtained by quantizing this algebra. Another typical model is the one employed in the current work (see, e.g., \cite{thiemann1998kinematical}). As aforementioned, this model considers the algebra formed by $\int\dd^3 y\sqrt{\delta(x,y)}\tilde\Psi(y)$, which leads to a fermionic Fock space as the Hilbert space for fermions. Our work develops this model from the following aspects.  At first, our work studies the Gauss constraint in details. We not only give the exact solutions to the Gauss constraint but also study the physical pictures resulting from these solutions. Second, for the Hamiltonian constraint of the fermion field, we propose a different  regularization strategy  than that in \cite{thiemann1998quantumV}. Additionally, the vertex Hilbert space is introduced, not only for removing the regulator in the regularized Hamiltonian constraint but also for fixing the issues on the adjoint of $\widehat{H(\delta)}$, so that a densely defined symmetric Hamiltonian constraint operator can be obtained.

This paper is organized as follows. In Sec. \ref{sec:two} the classical theory of gravity coupled to fermion field is introduced briefly. In Sec. \ref{sec:three} we introduce some basic notions of the kinematical Hilbert space of pure gravity and revisit the construction of the fermion kinematical Hilbert space. In Sec. \ref{sec:four}, the Gauss constraint and the Hamiltonian constraint are regularized and quantized, where the adapted vertex Hilbert space is introduced and some physical results are discussed. Finally, in \ref{sec:five}, we summarize the remarkable results and propose some outlooks for further works. 

\section{Classical theory of gravity  coupled to Fermion}\label{sec:two}
Let $\mathcal M$ denote the spacetime manifold, which is homeomorphism to $\mathbb R\times \Sigma$ with $\Sigma$ being the spatial manifold. Given a 4-dimensional vector space $V$, let $\eta_{IJ}$ be the Minkowski metric on it. A tetrad field $e_\mu^I$ gives  the metric $g_{\mu\nu}=\eta_{IJ}e^I_\mu e^J_\nu$ on $\mathcal M$. The curvature of $g_{\mu\nu}$ defines the Einstein-Hilbert action in terms of the tetrad fields,
\begin{equation}
S_H[e]=\frac{1}{2\kappa}\int_{\mathcal M}\dd^4 x R[e].
\end{equation}
where $\kappa=8\pi G$. 
 Let $\Gamma^I_{\mu J}$ denote the spin connection compatible with the tetrad so that
\begin{equation}
\dd e^I+\Gamma^I{}_J\wedge e^J=0.
\end{equation}
The model of gravity coupled to the fermion field is described by the action
\begin{equation}\label{eq:action}
\begin{aligned}
S[e,\Psi]=S_H[e]-\frac{i}{2}\int_{\mathcal M}\dd^4 x e\, \left(\overline\Psi\gamma^I e^\mu_I \nabla_\mu \Psi-c.c\right)
\end{aligned}
\end{equation}
where $\gamma^I$ denotes the gamma matrices satisfying $\gamma^I\gamma^J+\gamma^J\gamma^I=2\eta^{IJ}\mathbbm{1},$ and the covariant derivative of $\Psi$ is
\begin{equation}
\nabla_\mu\Psi=\partial_\mu\Psi-\frac{1}{4}\Gamma^I_{\mu J}\gamma^I\gamma^J\Psi. 
\end{equation}

Performing the 3+1-decomposition and doing canonical transformation \cite{thiemann2007modern}, we get the gravitational canonical pair $(A_a^i,E^a_i)$. The Poisson brackets between them are
\begin{equation}
\{A_a^i(x),E^b_j(y)\}=\kappa\beta\delta_a^b\delta^i_j\delta(x,y),
\end{equation}
where $\beta$ is the Barbero-Immirzi parameter. For the fermion field, we split the Dirac fermion $\Psi$ into its chiral components and
follow the argument in \cite{thiemann1998quantumV} to introduce the half densities on $\Sigma$,
\begin{equation}
\xi:=\sqrt[4]{q}\Psi_-,\ \nu=\sqrt[4]{q}\Psi_+
\end{equation}
with $\Psi_\pm=\frac{1\pm \gamma^5}{2}\Psi$. Detailed Hamiltonian analysis (see Appendix \ref{app:hamiltoniananalysisFermion}) tells that the conjugate moments to $\xi$ and $\nu$ are their complex conjugates, and the anti-Poisson brackets are
\begin{equation}\label{eq:symplecticFermion}
\begin{aligned}
\{\xi_A(x),\xi_B^\dagger(y)\}_+&=-i\delta_{AB}\delta(x,y),\\
\{\nu_A(x),\nu_B^\dagger(y)\}_+&=-i\delta_{AB}\delta(x,y).
\end{aligned}
\end{equation}
for all $A,B=\pm 1/2$.

The dynamics of this model is encoded in the Gauss constraint $G_m$, the diffeomorphism constraint $H_a$ and the Hamiltonian constraint $H$, which are 
\begin{equation}
\begin{aligned}
G_m=&\frac{1}{\kappa\beta}D_a E^a_l+\frac{1}{2} (\xi^\dagger\sigma_m\xi+\nu^\dagger\sigma_m\nu),\\
H_a=&\frac{1}{\kappa\beta} E^b_i F^i_{ab}+ \frac{i}{2}\Big\{\xi^\dagger D_a\xi-(D_a\xi)^\dagger\xi+\nu^\dagger D_a\nu-(D_a\nu)^\dagger\nu \Big\}+\beta K_a^m G_m,\\
H=&H_G+ \frac{1}{2\sqrt{q}}\Big[i(\xi^\dagger E_i^a\sigma^i D_a\xi-(D_a\xi)^\dagger E_i^a\sigma^i\xi)-\beta E^a_i K_a^i \xi^\dagger\xi-\frac{1}{\beta}(1+\beta^2)D_a E^a_i \xi^\dagger\sigma^i\xi-\beta E^a_iD_a\Big( \xi^\dagger\sigma^i\xi \Big)\\
&-i(\nu^\dagger E_i^a\sigma^iD_a\nu-(D_a\nu)^\dagger E_i^a\sigma^i\nu)+\beta E^a_i K_a^i \nu^\dagger \nu-\frac{1}{\beta}(1+\beta^2)D_a E^a_i\nu^\dagger \sigma^i\nu-\beta\frac{1}{\sqrt{q}} E^a_iD_a\Big(\nu^\dagger\sigma^i\nu\Big)\Big].
\end{aligned}
\end{equation}
Here, $H_G$ denotes the scalar constraint of pure gravity,
\begin{equation}
H_G=\frac{1}{2\kappa\sqrt{q}} E^a_i E^b_j\left(F_{ab}^m\epsilon_m{}^{ij}-2(1+\beta^2)K_{[a}^iK_{b]}^j\right).
\end{equation}

\section{Loop quantization of  the theory: kinematics}\label{sec:three}

\subsection{the kinematical Hilbert space of pure gravity}\label{sec:HG}
In LQG, besides a fixed differentiability class $C^m$ with $m\geq 1$,  a semianalytic structure on $\Sigma$ is also necessary \cite{lewandowski2005uniqueness}.  
 Then all local maps, differmorphisms, submanifold and function thereon are assumed to be $C^m$ and semianalytic. Particularly, an edge is a semianalytic curve embedded in $\Sigma$. A graph is a collection of edges $\{e_1,\cdots,e_n\}$ where these $e_k$ intersect each other at most at the ending points. 
Given a graph $\gamma\subset \Sigma$, let $E(\gamma)$ denote the set of its edges and  $V(\gamma)$, its vertices. The number of elements in $E(\gamma)$ ($V(\gamma)$ respectively) is denoted by $|E(\gamma)|$ ($|V(\gamma)|$ respectively).   
A cylindrical function $\Psi$ of the Ashtekar connection $A$  is a function that can be written in the form
\begin{equation}\label{eq:representationcyl}
\Psi(A)=\psi_\gamma(h_{e_1}(A),\cdots,h_{e_n}(A))
\end{equation}
where $\psi_\gamma:\sut^{|E(\gamma)|}\to \mathbb C$ is a complex function on $\sut^{|E(\gamma)|}$ and
$h_e(A)\in\sut$ is the parallel transport along an edge $e$ with respect to a given connection $A$,
\begin{equation}
h_e(A)=\mathcal P\exp\left(-\int_e A\right)=1+\sum_{n=1}^\infty(-1)^n\int_0^1\dd t_n\int_0^{t_n}\dd t_{n-1}\cdots\int_0^{t_2}\dd t_1A(t_1)\cdots A(t_n).
\end{equation}
Given a cylindrical function $\Psi$ with respect to a graph $\gamma$, it can always be rewritten via another graph $\gamma'\supset\gamma$. Therefore, for two cylindrical functions $\Psi^{(1)}$ and $\Psi^{(2)}$ with respect to graphs $\gamma_1$ and $\gamma_2$ respectively, one can always find another graph $\gamma_3$ with $\gamma_3\supset \gamma_1,\gamma_3\supset \gamma_2$, and rewrite $\Psi^{(1)}$ and  $\Psi^{(2)}$ by some functions $\psi_{\gamma_3}^{(1)}$ and $\psi_{\gamma_3}^{(2)}$ respectively on $\sut^{|E(\gamma_3)|}$. Then the inner product of $\Psi^{(1)}$ and $\Psi^{(2)}$ is
\begin{equation}\label{eq:innerproduct}
\langle\Psi^{(1)}|\Psi^{(2)}\rangle=\int_{\sut^{n}}\dd\mu_H(g)\overline{\psi_{\gamma_3}^{(1)}(g_1,\cdots,g_n)}\psi_{\gamma_3}^{(2)}(g_1,\cdots,g_n)
\end{equation}
where $n=|E(\gamma_3)|$ and $\dd\mu_H$ is the Haar measure on $\sut^n$.  Eq. \eqref{eq:innerproduct} defines a measure $\dd\mu_{\rm AL}$ on the quantum configuration space $\mathcal A$. Thus \eqref{eq:innerproduct} is always rewritten as
\begin{equation}\label{eq:innerproduct2}
\langle\Psi^{(1)}|\Psi^{(2)}\rangle=\int_{\mathcal A}\dd\mu_{\rm AL}(A)\,\overline{\Psi^{(1)}(A)}\Psi^{(2)}(A)
\end{equation}
The space of cylindrical function is denoted by $\cyl$. The Hilbert space $\mathcal H_G$ of the pure gravity is the completion of $\cyl$ with the inner product define in \eqref{eq:innerproduct2}. 

Given a graph $\gamma$, the space of cylindrical functions with respect to $\gamma$ is denoted by $\cyl_\gamma$. The Cauchy completion of $\cyl_\gamma$ with respect to \eqref{eq:innerproduct} is denoted by $\mathcal H_{G,\gamma}$. As shown in \cite{ashtekar2004back},  $\mathcal H_{G,\gamma}$ admits the spin network decomposition 
\begin{equation}
\mathcal H_{G,\gamma}=  {\bigoplus_{\vec j,\vec l}}\mathcal H_{G,\gamma}^{(\vec j,\vec l)}
\end{equation}
where $\vec j=\{j_1,j_2,\cdots,j_{|E(\gamma)|}\}$ assigns to each edge of $\gamma$ an irreducible representation of $\sut$, and $\vec l=\{l_1,\cdots,l_{|V(\gamma)|}\}$, to each vertex of $\gamma$ an irreducible representation. Consider the assignments $\vec j'$ such that each representation is non-trivial. Besides, let $\vec l'$ denote assignments of irreducible representations to vertices of $\gamma$ which are non-trivial at each spurious vertex of $\gamma$, where a vertex $v$ is spurious if it is bivalent, and if the edges $e_i$ and $e_{i+1}$ which meet at $v$ are such that $e_i\circ e_{i+1}$ is itself an semianalytic (i.e., $v$ just serves to split and edge). Then, we define $\mathcal H_{G,\gamma}',$ as
\begin{equation}\label{eq:decompositionG}
\mathcal H_{G,\gamma}'=  {\bigoplus_{\vec j',\vec l'}}\mathcal H_{G,\gamma}^{(\vec j',\vec l')}.
\end{equation}  
Thanks to $\mathcal H_{G,\gamma}'$, the space $\cyl$ can be decomposed as\footnote{For elements in an infinite direct sum, we require that all but finitely many components are zero.} 
\begin{equation}
\cyl=\bigoplus_\gamma \mathcal H_{G,\gamma}'\oplus \mathbb C.
\end{equation}

The multiplication operator $D^\iota_{ab}(h_e)$ acts on a cylindrical function $\Psi(A)=\psi_{\gamma}(h_{e_1}(A),\cdots,h_{e_n}(A))$ as
\begin{equation}
(D^\iota_{ab}(h_e)\Psi)(A)=D^\iota_{ab}(h_e(A))\psi_\gamma(h_{e_1}(A),\cdots,h_{e_n}(A)),
\end{equation}
 where $D^\iota_{ab}(h_e(A))$ denotes the Wigner-D matrix of $h_e(A)\in \sut$. Moreover, the derivative operators $\hat J_i^{v,e}$, for $i=1,2,3$ and $v$ being source point $s_e$ of $e$  or the target point $t_e$ of $e$, act  on $\Psi$ as
\begin{equation}
(\hat J_j^{v,e}\Psi)(A)=\left\{
\begin{aligned}
&i\left.\frac{\dd}{\dd t}\right|_{t=0}\psi_{\gamma}(h_{e_1},\cdots,h_ee^{t\tau_j},\cdots,h_{e_n}),\  v=s_e,\\
&i\left.\frac{\dd}{\dd t}\right|_{t=0}\psi_{\gamma}(h_{e_1},\cdots,e^{-t\tau_j}h_e,\cdots,h_{e_n}),\  v=t_e,
\end{aligned}
\right.
\end{equation}
where $\tau_j=-i \sigma_j/2$ with $\sigma_j$ being the Pauli matrices.  With the operator $\hat J_j^{v,e}$, we can define an operator $\hat J_j^{x,[e]}$ such that
\begin{equation}
\hat J_j^{x,[e]}\Psi=\sum_{e'\in [e]}\hat J^{x,e'}_j\Psi, \forall \Psi\in\cyl,
\end{equation}
where  $[e]$ is a maximal family of curves beginning at $x\in\Sigma$ such that each two curves overlap on a connected initial segment containing $x$.

\subsection{loop quantization of the fermion field}\label{sec:HF}
From now on, we will only focus on the single Weyl component $\xi$. However, everything works similarly for the other chiral component $\nu$. 
To quantize the fermion field, we follow \cite{thiemann1998quantumV} to use the modified symplectic structure
\begin{equation}\label{eq:modifiedpoisson}
\{\theta_A(x),\theta_B^\dagger(y)\}_+=-i\delta_{AB}\delta_{x,y},\ A,B=\pm\frac{1}{2}
\end{equation}
where,   {in comparison with} \eqref{eq:symplecticFermion}, the Dirac delta $\delta(x,y)$ is changed to the Kronecker delta $\delta_{x,y}$. 
This change   {results from} the following canonical transformation from $\xi$ to $\theta$:
\begin{equation}\label{eq:relation}
\begin{aligned}
\theta(x)&=\int_\Sigma\dd^3 y \sqrt{\delta(x,y)}\xi(y),\\
\xi(x)&=\sum_{y\in\Sigma}\sqrt{\delta(x,y)}\theta(y).
\end{aligned}
\end{equation}
To prove the relation \eqref{eq:relation}, one used the function $f_\epsilon(x,y):=\chi_\epsilon(x,y)/\epsilon^3$ to regularize the Dirac delta function \cite{thiemann1998quantumV}, where $\chi_\epsilon$ is 
$$
\chi_\epsilon(x,y):=\left\{
\begin{aligned}
1,\ & \sum_{a=1}^3(x^a)^2+(y^a)^2\leq\left(\frac{\epsilon}{2}\right)^2,\\
0,\ & \text{otherwise}.
\end{aligned}
\right.
$$
According to \eqref{eq:relation}, $\xi(x)$ will be singular for regular $\theta(x)$. This scenario contradicts that $\xi(x)$ is smooth as a classical field. To have a consistent understanding of this formulation, one interprets the singular fields $\xi(x)$ as defining the quantum configuration space of the fermion field so that $\theta(x)$ is a regular-field coordinate of this quantum configuration space. 

As in pure LQG, the quantization starts by introducing the cylindrical functions. Before doing so, we will first introduce a convenient field $\zeta_A$ as
\begin{equation}
\zeta_A(x)=\frac{1}{\sqrt{\hbar}}\theta_A(x),\ A=\pm\frac{1}{2}.
\end{equation}

\subsubsection{the cylindrical functions of the fermion field}

A fermionic graph $\gamma_F$ is  a finite subset of $\Sigma$ with $|\gamma_F|$ elements. Elements in $\gamma_F$ are  called fermionic vertices. An orientation of $\gamma_F$ is   a  surjection  $n\mapsto v_F^{(n)}\in \gamma_F$ with $1\leq n\leq |\gamma_F|$. The surjection endows the elements in $\gamma_F$ with an order.
Given  an oriented graph  $\gamma_F$, we have a family of Grassmann numbers  $\{\zeta^\dagger_{\pm \frac12}(v_F)\}_{v_F\in \gamma_F}$, which will be renamed to $\rho^\dagger_n$ with $1\leq n\leq 2|\gamma_F|$ such that
\begin{equation}\label{eq:zetafieldtograssmann}
\begin{aligned}
  \rho_{2i-1}^\dagger\equiv \zeta_\frac{1}{2}^\dagger(v_F^{(i)}),\ \rho_{2i}^\dagger\equiv \zeta_{-\frac{1}{2}}^\dagger(v_F^{(i)}),\ 1\leq i\leq |\gamma_F|.  
\end{aligned}
\end{equation}
Then a function $\Psi$ of $\zeta^\dagger$ with respect to $\gamma_F$ takes the general form 
\begin{equation}\label{eq:cylindricalfunctionfermion}
\Psi(\zeta^\dagger)=f_0+\sum_{n=1}^{2|\gamma_F|}\sum_{1\leq i_1<i_2<\cdots <i_n\leq   {2|\gamma_F|}}f_{i_1\cdots i_n} \rho^\dagger_{i_1}\rho^\dagger_{i_2}\cdots\rho^\dagger_{i_n},
\end{equation}
where $f_0$ and $f_{i_1\cdots i_n}$ are complex numbers. Functions taking the form \eqref{eq:cylindricalfunctionfermion} are called the (fermionic) cylindrical functions of $\zeta^\dagger$. The space of cylindrical functions of $\zeta^\dagger$ will be denoted by $\cylf$. 

According to \eqref{eq:cylindricalfunctionfermion}, each cylindrical function with respect to $\gamma_F$ can be identical with a vector $$f_{\gamma_F}=\left(f_0,\{f_{i_1\cdots i_n}\}_{1\leq i_1<i_2<\cdots <i_n\leq   {2|\gamma_F|}}\right)\in\mathbb C^{2^{2|\gamma_F|}}.$$
Moreover, for a cylindrical function $\Psi$ expressed via $f_{\gamma_F}\in\mathbb C^{2^{2|\gamma_F|}}$, there always exists a lager graph $\gamma_F'\supset\gamma_F$ such that $\Psi$ is rewritten via  some $f'_{\gamma_F'}\in\mathbb C^{2^{2|\gamma_F'|}}$. Thus, given two functions $\Psi_1,\Psi_2\in\cylf$  on $\gamma_F{}_1$ and  $\gamma_F{}_2$ respectively, we can find  another graph $\gamma_F'$ containing both $\gamma_F{}_1$ and $\gamma_F{}_2$ to rewrite $\Psi_1$  and  $\Psi_2$ with $f_{\gamma_F'}^{(1)},f_{\gamma_F'}^{(2)}\in\mathbb C^{2^{2|\gamma_F'|}}$. Then  the inner  product of $\Psi_1$ and $\Psi_2$ is given by
\begin{equation}\label{eq:innerproductfermion}
\langle\Psi_1,\Psi_2\rangle=\int \dd\mu(\rho_1\rho_1^\dagger)\dd\mu(\rho_2\rho_2^\dagger)\cdots \dd\mu(\rho_{2|\gamma_F'|}\rho_{2|\gamma_F|'}^\dagger)\Psi_1^\dagger\Psi_2
\end{equation}
with $\dd\mu(\rho_n\rho_n^\dagger)=\dd\rho_n^\dagger\dd\rho_ne^{\rho_n\rho_n^\dagger}$. By applying  \eqref{eq:innerproductfermion}, one can verify
\begin{equation}
\langle\Psi_1,\Psi_2\rangle=(f_{\gamma_F'}^{(1)})^\dagger f_{\gamma_F'}^{(2)}. 
\end{equation}
Moreover, even though the graphs $\gamma_F'$ containing both $\gamma_F{}_1$ and $\gamma_F{}_2$ are not unique,
Eq. \eqref{eq:innerproductfermion} is independent of the choice of $\gamma_F'$ since $
\int\dd\mu(\rho_n\rho_n^\dagger)=1$. Indeed, Eq. \eqref{eq:innerproductfermion} defines a measure $\dd\mu_H(\zeta^\dagger\zeta)$ on $\cylf^\dagger\otimes \cylf$, and will be rewritten as
\begin{equation}\label{eq:fermionhaarmeasure}
\langle\Psi_1,\Psi_2\rangle=\int \dd\mu_H(\zeta\zeta^\dagger) \Psi_1^\dagger \Psi_2. 
\end{equation}
The fermion Hilbert space $\mathcal H_F$ is the completion of $\cylf$ with the inner product defined by \eqref{eq:fermionhaarmeasure}, i.e.,
\begin{equation}
\mathcal H_F=\overline{\cylf}. 
\end{equation}

On $\cylf$, a type of operators are the multiplication operators $\widehat{\zeta_{v_F,A}^\dagger}$ given by
\begin{equation}
(\widehat{\zeta_{v_F,A}^\dagger}\Psi)(\zeta^\dagger)=\zeta_A^\dagger(v_F)\Psi(\zeta^\dagger),\ \forall \Psi\in\cylf.  
\end{equation}
Another type of operators are the  derivative operators $\hat\zeta_{v_F,A}$, acting on $\Psi\in \cylf$ as
\begin{equation}
(\hat\zeta_{v_F,A}\Psi)(\zeta^\dagger)=\left(\frac{\partial}{\partial \zeta_A^\dagger(v_F)}\Psi\right)(\zeta^\dagger),\ \forall \Psi\in\cylf.  
\end{equation}
It is easy to verify that $\widehat{\zeta_{v_F,A}^\dagger}$ and $\hat\zeta_{v_F,A}$ are adjoint to each other, i.e.,
\begin{equation}
\widehat{\zeta_{v_F,A}^\dagger}=\hat\zeta_{v_F,A}^\dagger,
\end{equation}
 which realizes the real condition. Moreover, one has
\begin{equation}
[  {\hat\zeta}_{v_F,A},  {\hat\zeta}^\dagger_{v_F',B}]_+=  {\hat\zeta}_{v_F,A},  {\hat\zeta}^\dagger_{v_F',B}+  {\hat\zeta}^\dagger_{v_F',B}  {\hat\zeta}_{v_F,A}=\delta_{AB}\delta_{v_Fv_F'},
\end{equation} 
which implements the Poisson brackets \eqref{eq:modifiedpoisson} by  defining
\begin{equation}
\hat\theta_{A}(v_F)=\sqrt{\hbar}\, \hat\zeta_{v_F,A}. 
\end{equation}

\subsubsection{the spin network states of fermion  field }\label{sec:hilberttoeachx}
Given a graph $\gamma_F$, the space  of the fermionic cylindrical functions with respect to $\gamma_F$ is a finite-dimensional Hilbert space, denoted by $\mathcal H_{  {F,}\gamma_F}$.
Considering a graph $\gamma_F=\{v_F\}$ which is compronent of a single vertex, one has the the space $\mathcal H_{  {F,}\{v_F\}}\equiv\mathcal H_{v_F}$  consisting of  functions 
\begin{equation}
\Psi(\zeta^\dagger)=a_{00}+a_{10}\zeta_\frac12^\dagger(v_F)+a_{01}\zeta_{-\frac12}^\dagger(v_F)+a_{11}\zeta_{\frac12}^\dagger(v_F)\zeta_{-\frac12}^\dagger(v_F)
\end{equation}
The inner product of $\Psi^{(i)}(\zeta^\dagger)=a_{00}^{(i)}+a_{10}^{(i)}\zeta_{\frac12}^\dagger(v_F)+a_{01}^{(i)}\zeta_{-\frac12}^\dagger(v_F)+a_{11}^{(i)}\zeta_{\frac12}^\dagger(v_F)\zeta_{-\frac12}^\dagger(v_F)$ with  $i=1,2$ is
\begin{equation}
\langle\Psi^{(1)},\Psi^{(2)}\rangle=\sum_{i,j\in\{0,1\}}(a_{ij}^{(1)})^*a_{ij}^{(2)}
\end{equation}

For convenience, we introduce the Dirac bra-ket notation $|i,j\rangle_{v_F}$ ($i,j=0,1$) to denote the state  $\Psi_{ij}$, where $\Psi_{ij}$ are the states given by $\Psi_{00}(\zeta^\dagger)=1$, $\Psi_{10}(\zeta^\dagger)=\zeta_\frac12^\dagger(v_F)$, $\Psi_{01}(\zeta^\dagger)=\zeta_{-\frac12}^\dagger(v_F)$ and $\Psi_{11}(\zeta^\dagger)=\zeta_\frac12^\dagger(v_F)\zeta_{-\frac12}^\dagger(v_F)$. Then the states $|i,j\rangle_{v_F}$ form an orthonormal basis of $\mathcal H_{v_F}$, i.e.,
\begin{equation}
{}_{v_F}\langle i_1,j_1|i_2,j_2\rangle_{v_F}=\delta_{i_1i_2}\delta_{j_1j_2}. 
\end{equation}
The action of $\hat\zeta_{v_F,A}$ and $\hat\zeta_{v_F,A}^\dagger$ for $A=\pm \frac12$  on $\mathcal H_{v_F}$ reads
\begin{equation}
\begin{array}{lll}
\hat\zeta_{v_F,\frac12}^\dagger|0,i_2\rangle_{v_F}=|1,i_2\rangle_{v_F},& \hat\zeta_{v_F,\frac12}^\dagger|1,i_2\rangle_{v_F}=0,& \forall i_2=0,1,\\
\hat\zeta_{v_F,\frac12}|0,i_2\rangle_{v_F}=0,& \hat\zeta_{v_F,\frac12}|1,i_2\rangle_{v_F}=|0,i_2\rangle_{v_F},& \forall i_2=0,1,\\
\hat\zeta_{v_F,-\frac12}^\dagger|i_1,0\rangle_{v_F}=(-1)^{i_1}|i_1,1\rangle_{v_F},& \hat\zeta_{v_F,-\frac12}^\dagger|i_1,1\rangle_{v_F}=0,&\forall i_1=0,1,\\
\hat\zeta_{v_F,-\frac12}|i_1,0\rangle_{v_F}=0,&\hat\zeta_{v_F,-\frac12}|i_1,1\rangle_{v_F}=(-1)^{i_1}|i_1,0\rangle_{v_F},& \forall i_1=0,1.
\end{array}
\end{equation}

For the general case where $\gamma_F\subset\Sigma$ consists of more than one vertices, we first associate to each $v_F\in\gamma_F$ the Hilbert space $\mathcal H_{v_F}$. Then, fixing an orientation of $\gamma_F$, one has the tensor product space
\begin{equation}
\mathcal H_{F,\gamma_F}=\mathcal H_{v_F^{(1)}}\otimes \mathcal H_{v_F^{(2)}}\otimes\cdots\otimes \mathcal H_{v_F^{|\gamma_F|}}. 
\end{equation}
An orthonormal basis of $\mathcal H_{F,\gamma_F}$ is composed of the vectors
\begin{equation}\label{eq:basis0}
|i_1,i_2,\cdots,i_{2|\gamma_F|}\rangle:=|i_1,i_2\rangle_{v_F^{(1)}}\otimes |i_3,i_4\rangle_{v_F^{(2)}}\otimes\cdots\otimes |i_{2|\gamma_F|-1},i_{2|\gamma_F|}\rangle_{v_F^{(|\gamma_F|)}}
\end{equation}
with $i_k\in\{0,1\}$ for all $1\leq k\leq 2|\gamma_F|$. Indeed, the vector $|i_1,i_2,\cdots,i_{2|\gamma_F|}\rangle$ refers to the cylindrical function $\Psi_{\vec i}=\rho_{i_1}^\dagger\rho_{i_2}^\dagger\cdots \rho_{i_n}^\dagger$ with respect to $\gamma_F$.
It is worth noting that the definition of $\mathcal H_{F,\gamma_F}$ depends on the orientation of $\gamma_F$, since $\zeta_A^{  {\dagger}}(v_F)$ are Grassmann numbers.  This fact can be illustrated more explicitly with the following examples.
Consider another orientation $n\to\tilde v_F^{(n)}$ of $\gamma_F$ such that
\begin{equation}\label{eq:neworientation}
\tilde v_F^{(1)}=v_F^{(2)},\tilde v_F^{(2)}=v_F^{(1)},  \tilde v_F^{(k)}=v_F^{(k)},\ \forall k\geq 3.
\end{equation} 
Then, under the new orientation, we have the tensor-product Hilbert space $\tilde{\mathcal H}_{F,\gamma_F}$ possessing  the basis
\begin{equation}\label{eq:tildebasis}
\widetilde{|i_1,i_2,\cdots,i_{2|\gamma_F|}\rangle}:=|i_1,i_2\rangle_{\tilde v_F^{(1)}}\otimes |i_3,i_4\rangle_{\tilde v_F^{(2)}}\otimes\cdots\otimes |i_{2|\gamma_F|-1},i_{2|\gamma_F|}\rangle_{\tilde v_F^{(\gamma_F)}}.
\end{equation}
By definition, $\widetilde{|i_1,i_2,\cdots,i_{2|\gamma_F|}\rangle}$ referes to the cylindrical funciton
\begin{equation}\label{eq:cylindricalfunciton}
\tilde \Psi_{\vec i}=\rho_{i_3}^\dagger\rho_{i_4}^\dagger\rho_{i_1}^\dagger\rho_{i_2}^\dagger\cdots \rho_{i_n}^\dagger=(-1)^{(i_1+i_2)(i_3+i_4)}\Psi_{\vec i'},
\end{equation}
where $\vec i'=\{i_3,i_4,i_1,i_2,\cdots,i_{2|\gamma_F|}\}$. Eq. \eqref{eq:cylindricalfunciton} gives the equivalence relation between $\mathcal H_{F,\gamma_F}$ and $\tilde{\mathcal H}_{F,\gamma_F}$
\begin{equation}\label{eq:identitydifferentori}
\widetilde{|i_1,i_2,\cdots,i_{2|\gamma_F|}\rangle}= (-1)^{(i_1+i_2)(i_3+i_4)}|i_3,i_4,i_1,i_2,\cdots,i_{2|\gamma_F|}\rangle. 
\end{equation}
This equivalence relation can  be analogously defined between the tensor product spaces with different orientations. The fermionic Hilbert space with respect to $\gamma_F$ is actually   {the} space of equivalence classes associated with  this equivalence relation. 

Indeed, the extra sign in \eqref{eq:identitydifferentori} can be systematically obtained by introducing the notion of graded objects. One can refer to Appendix \ref{app:graded} and the reference therein for more details on the this notion.
In our work, the Hilbert spaces $\mathcal H_{v_F}$ are graded. The degree $\grade(i_1,i_2)$ of each $|i_1,i_2\rangle_{v_F}$  is
\begin{equation}
\grade(i_1,i_2)=i_1+i_2\mod 2.
\end{equation}
The operator algebra on $\mathcal H_{v_F}$ is also graded. By definition, the degrees of the operators $\hat\zeta_{v_F,A}$ and $\hat\zeta_{v_F,A}^\dagger$ are
\begin{equation}
\grade(\hat\zeta_{v_F,A})=1=\grade(\hat\zeta_{v_F,A}^\dagger).
\end{equation}
A general principle to deal with these graded objects is that, whenever we swap two items, an additional sign appears by the rule $xy=(-1)^{\grade(x)\grade(y)}y x$. 
Following this rule, we can obtain the identity \eqref{eq:identitydifferentori} manifestly.

Given a vertex $v_F$, let $\mathcal H_{v_F}^{\rm irr}$  be the subspace of $\mathcal H_{v_F}$ spanned by $|0,1\rangle_{v_F}$,  $|1,0\rangle_{v_F}$ and $|1,1\rangle_{v_F}$. Then the Hilbert space $\mathcal H_{F,\gamma_F}^{\rm irr}$ with respect to $\gamma_F$ is defined as
\begin{equation}
\mathcal H_{F,\gamma_F}^{\rm irr}=\bigotimes_{v_F\in\gamma_F}\mathcal H_{v_F}^{\rm irr},
\end{equation}
such that $\cyl_F$ can be  decomposed as
\begin{equation}
\cyl_F=\bigoplus_{\gamma_F}\mathcal H_{F,\gamma_F}^{\rm irr}\oplus \mathbb C.
\end{equation}

The kinematical Hilbert space $\mathcal H$ of the entire system is the tensor product of $\mathcal H_G$ and $\mathcal H_F$, i.e.,
\begin{equation}
\mathcal H=\overline{\mathcal H_G\otimes\mathcal H_F}.
\end{equation}
A densely subspace $\cylt$ of $\mathcal H$ is 
\begin{equation}
\cylt=\cyl\otimes\cylf.
\end{equation}
The states in $\cylt$ will be called  the cylindrical states. To obtain a cylindrical  state, one needs a graph $\gamma=\gamma_G\cup\gamma_F$,  where the gravitational  graph $\gamma_G$ is constituted of edges and their ending points as vertices, and the fermionic graph $\gamma_F$ contains only vertices . To define a state with respect to $\gamma$, besides the data for a (gauge variant) LQG spin network state, one also needs to assign to each fermionic vertex $v_F$ a state  $|i_1(v_F),i_2(v_F)\rangle_{v_F}$ with $i_1(v_F),i_2(v_F)\in\{0,1\}$. In principle, a fermionic vertex $v_F$ can be located anywhere, regardless of the given gravitational graph $\gamma_G$. However, if $v_F$ is chosen as point in $e\in E(\gamma_G)$ but $v_F\notin V(\gamma_G)$, we can always split $e$ at $v_F$ to define a new graph $\tilde\gamma_G$. Then $v_F$ becomes a vertex of $\tilde\gamma_G$.  Moreover, because of $\gamma_G\subset \tilde\gamma_G$, every cylindrical function with respect to $\gamma_G$ can be rewritten by using $\tilde\gamma_G$. Thus, it is sufficient to consider those graphs $\gamma=\gamma_G\cup\gamma_F$ where each fermionic vertex $v_F$ satisfies either $v_F\in V(\gamma_G)$ or $v_F\notin\gamma_G$. Then, in $V(\gamma_G)$, there could be bivalent vertices which is a fermionic vertex. These vertices serve to split edge. Thus, by  \eqref{eq:decompositionG}, they are just spurious vertices. However, in contrast to \eqref{eq:decompositionG}, we can put trivial SU(2) representation at these vertices to decompose the Hilbert space with respect to graphs. Moreover precisely, let $\Gamma_o$ be the set of graphs $\gamma=\gamma_G\cup\gamma_F$ where each fermionic vertex $v_F$ satisfies either $v_F\in V(\gamma_G)$ or $v_F\notin\gamma_G$.  Consider the assignments $\vec j'$ to $E(\gamma)$ such that each representation is nontrivial. Besides, let $\vec l'$ denote assignments of irreducible representations to vertices of $\gamma$ which are non-trivial at each fake vertex of $\gamma_G$ where a vertex $v\in V(\gamma_G)$ is fake if $v\notin \gamma_F$ and it is spurious as defined in \eqref{eq:decompositionG}. Then, we define
\begin{equation}
\mathcal H_{G,\gamma_G}^{\rm irr}:=  {\bigoplus_{\vec j',\vec l'}}\mathcal H_{G,\gamma_G}^{(\vec j',\vec l')},
\end{equation}
which gives us the Hilbert space $\mathcal H_\gamma^{\rm irr}$ as
\begin{equation}
\mathcal H_\gamma^{\rm irr}=\mathcal H_{G,\gamma_G}^{\rm irr}\otimes\mathcal H_{F,\gamma_F}^{\rm irr}. 
\end{equation}
Then we have the decomposition
\begin{equation}\label{eq:decompsitionfull}
\cylt=\bigoplus_{\gamma\in\Gamma_o}\mathcal H_\gamma^{\rm irr}\oplus\mathbb C.
\end{equation}

\section{The constraint operators for gravity coupled to fermion field}\label{sec:four}
\subsection{the Gauss constraint}\label{sec:gaussian}
Classically, the Gauss constraint $G[\lambda]$ reads
\begin{equation}
G[\lambda]=\int_\Sigma\dd^3 x\lambda^m\left(\frac{1}{\kappa\beta}D_aE^a_m+\frac{1}{2}\xi^\dagger\sigma_m\xi\right).
\end{equation}
It is straightfoward to quantize it as the operator
\begin{equation}
\widehat{G[\lambda]}=\sum_{v}\lambda^m(v)\hat G_{v,m}
\end{equation}
with
\begin{equation}\label{eq:Gaussian}
\hat G_{v,m}=\hbar\sum_{[e]}\hat J_m^{v,[e]}+\hbar\,\hat \zeta_{v,A}^\dagger \frac{(\sigma_m)_{AB}}{2}\hat\zeta_{v,B}
\end{equation}
Let us use $\hat{\mathcal J}_{v,m}$ to denote the second term in \eqref{eq:Gaussian}, namely
\begin{equation}
\hat{\mathcal J}_{v,m}=\hat \zeta_{v,A}^\dagger \frac{(\sigma_m)_{AB}}{2}\hat\zeta_{v,B}.
\end{equation}
On the fermionic Hilbert space $\mathcal H_v$ at vertex $v$, the action of  $\hat{\mathcal J}_{v,m}$ reads
\begin{equation}\label{eq:basis}
\begin{aligned}
\hat{\mathcal J}_{v,m}|0,0\rangle_v&=0,\ \hat{\mathcal J}_{v,m}|1,1\rangle_v=0\\
\hat{\mathcal J}_{v,m}\left(|1,0\rangle_v,|0,1\rangle_v\right)&=\left(|1,0\rangle_v,|0,1\rangle_v\right)\frac{\sigma_m}{2}.
\end{aligned}
\end{equation}
According to \eqref{eq:basis}, the operators $\hat{\mathcal J}_{v,m}$ for all $m=1,2,3$ behave as the angular moment operators. Thus, the operator $\hat{\mathcal J}_{v,m}$ generates an SU(2) action on $\mathcal H_v$ as 
\begin{equation}
u\triangleright|\phi\rangle_v=\begin{pmatrix}
|1,0\rangle_v,|0,1\rangle_v
\end{pmatrix} u
\begin{pmatrix}
\phi_{10}\\
\phi_{01}
\end{pmatrix}+\phi_{00}|0,0\rangle_v+\phi_{11}|1,1\rangle_v.
\end{equation} 
where $|\phi\rangle_v=\sum_{ij}\phi_{ij}|i,j\rangle_v$ and $u\in \sut$.
Therefore, $\mathcal H_v$ becomes a reducible representation space of SU(2). The 1-dimensional space spanned by either $|0,0\rangle_v$ or $|1,1\rangle_v$ is the trivial representation space, and the 2-dimensional space spanned by $|0,1\rangle_v$ and $|1,0\rangle_v$ is the $1/2$-representation space where $|0,1\rangle_v$ and $|1,0\rangle_v$ serve as the standard basis according to \eqref{eq:basis}. This fact leads to the decomposition 
\begin{equation}\label{eq:isometricSU2}
\mathcal H_v\equiv \mathcal H_0\oplus\mathcal H_0\oplus\mathcal H_{1/2},
\end{equation}
where $\mathcal H_j$ denote the $j$-representation space of SU(2). 

For a graph  $\gamma=\gamma_G\cup\gamma_F$,  a spin $j_e$ is assigned to the edge $e\subset \gamma_G$. Then at each vertex $v\in V(\gamma)$ there is the Hilbert space 
\begin{equation}
\mathcal H_v^{\rm tot}=\bigotimes_{e\text{ starts from } v}\mathcal H_{j_e}\otimes \bigotimes_{e' \text{ targets } v}\mathcal H_{j_{e'}}^*\otimes \mathcal H_v
\end{equation}
where $\mathcal H_j^*$ denotes the dual space of $\mathcal H_j$. On  $\mathcal H_v^{\rm tot}$, the infinitesimal  SU(2) action gives the Gauss constraint. Thus, the solution space to the Gauss constraint is
\begin{equation}
\mathcal H^{\rm Gau}=\bigotimes_{x\in V(\gamma)}{\rm Inv}\left(\mathcal H_v^{\rm tot}\right),
\end{equation}
where ${\rm Inv}\left(\mathcal H_v^{\rm tot}\right)\subset \mathcal H_v^{\rm tot}$  is the SU(2)-invariant subspace. 
To see ${\rm Inv}(\mathcal H_v^{\rm tot})$  more precisely, let us  assume all edges at $v$ are outgoing without  loss of generality. 
Then we have
\begin{equation}
\mathcal H_v^{\rm tot}=\bigotimes_{e \text{ at } v}\mathcal H_{j_e}\otimes  \mathcal H_v.
\end{equation}
Given an order of the edges at $v$, one can choose an orthonormal basis of  $\bigotimes\limits_{e \text{ at } v}\mathcal H_{j_e}$ composed  of vectors $|k_{2},k_{3},\cdots,k_{n},M\rangle$ satisfying
\begin{equation}
\begin{aligned}
\sum_{i=1}^3(L_i^{(l)})^2|k_{2},k_{3},\cdots,k_{n},M\rangle=&k_l(k_l+1)|k_{2},k_{3},\cdots,k_{n},M\rangle,\ \forall l=2,\cdots, n\\
L_3^{(n)}|k_{2},k_{3},\cdots,k_{n},M\rangle=&M|k_{2},k_{3},\cdots,k_{n},M\rangle,\\
L_1^{(n)}|k_{2},k_{3},\cdots,k_{n},M\rangle=&\sum_{s=\pm 1}\frac{1}{2}\sqrt{(k_n-s M)(k_n+s M+1)}|k_{2},k_{3},\cdots,k_{n},M+s\rangle\\
L_2^{(n)}|k_{2},k_{3},\cdots,k_{n},M\rangle=&\sum_{s=\pm 1}\frac{-i s}{2}\sqrt{(k_n-s M)(k_n+s M+1)}|k_{2},k_{3},\cdots,k_{n},M+s\rangle
\end{aligned}
\end{equation}
with $\hat L_i^{(l)}:=\sum_{k=1}^l\hat J_i^{v,e_k}.$ 
Let us define $\mathrm{Inv}\left(\mathcal H_G^{(v)}\right)\subset \bigotimes\limits_{e \text{ at } v}\mathcal H_{j_e}$ as the subspace spanned by $|k_{2},k_{3},\cdots,k_{n-1},0,0\rangle$ for all possible $k_2,k_3,\cdots,k_{n-1}$. One has 
\begin{equation}
L_i^{(n)}|k_{2},k_{3},\cdots,k_{n-1},0,0\rangle=0,\ \forall i=1,2,3.
\end{equation}
 Moreover, with the vectors $|k_{2},k_{3},\cdots,k_{n-1},1/2,M\rangle$, we define
 \begin{equation}\label{eq:gaugeinvtotal}
 |k_{2},k_{3},\cdots,k_{n-1}\rangle_{\rm tot}=  {\frac{1}{\sqrt{2}}}\left|k_{2},k_{3},\cdots,k_{n-1},\frac12,\frac12\right\rangle\otimes |0,1\rangle_v-  {\frac{1}{\sqrt{2}}}\left|k_{2},k_{3},\cdots,k_{n-1},\frac12,-\frac12\right\rangle\otimes |1,0\rangle_v.
 \end{equation}
 Then one  has  
 \begin{equation}
( L_i^{(n)}+\hat{\mathcal J}_{v,i})|k_{2},k_{3},\cdots,k_{n-1}\rangle_{\rm tot}=0,\ \forall i=1,2,3.
 \end{equation}
 Let $\mathcal H_{\rm inv}\subset \mathcal H_v^{\rm tot}$ denote the subspace spanned by $|k_{2},k_{3},\cdots,k_{n-1}\rangle_{\rm tot}$ for all possible $k_2,k_3,\cdots,k_{n-1}$. Then ${\rm Inv}(\mathcal H_v^{\rm tot})$ can be decomposed as 
 \begin{equation}\label{eq:decomposehxinv}
 {\rm Inv}(\mathcal H_v^{\rm tot})=\left(\mathrm{Inv}\left(\mathcal H_G^{(v)}\right)\otimes |0,0\rangle_v\right)\oplus \left(\mathrm{Inv}\left(\mathcal H_G^{(v)}\right)\otimes |1,1\rangle_v\right)\oplus \mathcal H_{\rm inv},
 \end{equation}
 where  $\mathrm{Inv}\left(\mathcal H_G^{(v)}\right)\otimes |i_1,i_2\rangle_v$ is the space composed of vectors $|\psi\rangle\otimes |i_1,i_2\rangle$ for all $|\psi\rangle\in \mathrm{Inv}\left(\mathcal H_G^{(v)}\right)$.
 
 Let $v$ be a $n$-valence gauge invariant fermionic vertex, where the $i$th edge $e_i$ is assigned to a spin $j_i$.
 According to the decomposition \eqref{eq:decomposehxinv}, the gauge invariant Hilbert space $ {\rm Inv}(\mathcal H_v^{\rm tot})$ contains   {the} subspace $\mathcal H_{\rm inv}$,  isometric to the gauge invariant Hilbert space of a $(n+1)$-valence pure-gravity vertex  where the $i$th edge for $1\leq i\leq n$  is assigned to  spin $j_i$, and  the $(n+1)$th, the spin $1/2$.
Then once we  consider  the volume operator at $v$, this  extra spin $1/2$ will also  have contribution.  
Since the extra spin $1/2$ originates from the fermion filed, one gets an intuitive picture that fermion field contributes to the volume of a vertex. Moreover, a $n$-valence vertex  in pure-LQG is always regarded as a polyhedron whose faces are dual to the edges. The flux operators $\hat J^{v,e}_i$ associated to each edge $e$ have the geometric interpretation of the area vector of the dual face.  Then the pure-LQG Gauss constraint is just the   {closure} condition $\sum_{e}\hat J^{v,e}_i=0$ ensuring that the faces can form a closed polyhedron. Now, the fermion field is involved. Then the Gauss constraint \eqref{eq:Gaussian} implies 
\begin{equation}\label{eq:closercondition}
\sum_{e}\hat J^{v,e}_i=-\hat{\mathcal J}_{v,i}
\end{equation}
where the right hand side does not vanish in general. Thus, the faces dual  to the edges could not  give a closed polyhedron for states in $\mathcal H_{\rm inv}$. By \eqref{eq:closercondition},  the area defect of this unclosed polyhedron is filled by $\hat{\mathcal J}_{v,i}$, i.e. the fermion spin at the vertex (see \cite{mansuroglu2021fermion}  for more details on the fermion spin).   A direct consequence of the above discussion is that the volume of a 3-valence vertex with fermion does not vanish any more for states in $\mathcal H_{\rm inv}$. Let $j_i$ with $i=1,2,3$ be the spins on the edges. Then the states in $\mathcal H_{\rm inv}$ are spanned by $|k\rangle_{\rm tot}\equiv |k\rangle$ with  $k=j_3\pm 1/2$. The action of the operator $\hat q_{123}$, the operator proportional to the square of the volume operator \cite{ashtekar2004back},  on $ |k\rangle$ reads
\begin{equation}
\begin{aligned}
\langle k|\hat q_{123}|k+1\rangle=&\frac{-i}{4\sqrt{(2k+1)(2k+3)}} \sqrt{(j_1-j_2+k+1)(-j_1+j_2+k+1)(j_1+j_2-k)(j_1+j_2+k+2)} \\
&\sqrt{(j_3-\frac12+k+1)(-j_3+\frac12+k+1)(j_3+\frac12-k)(j_3+\frac12+k+2)}.
\end{aligned}
\end{equation}
Then we  have
\begin{equation}
\begin{aligned}
\left\langle j_3-\frac12\middle|\hat q_{123}\middle|j_3+\frac12\right\rangle=&i\frac{1}{16}  \sqrt{(2 j_1+2 j_2-2 j_3+1)(2 j_1-2 j_2+2 j_3+1)} \\
&\sqrt{(-2 j_1+2 j_2+2 j_3+1)(2 j_1+2 j_2+2 j_3+3)}.
\end{aligned}
\end{equation}
Since  the associated Hilbert space is 2-dimensional, the whole Hilbert space is the  eigenspace of the volume operator $\frac{\kappa_0  {\beta} \Pl^{3/2}}{  {2\sqrt{2}}}\sqrt{|\hat q_{123}|}$
 with eigenvalue
\begin{equation}
V_v=\frac{\kappa_0  {\beta} \Pl^{3/2}}{  {2\sqrt{2}}}\sqrt{\left|\left\langle j_3-\frac12\middle|\hat q_{123}\middle|j_3+\frac12\right\rangle\right|}.
\end{equation}

\subsection{The Hamiltonian constraint}\label{sec:Hoperators}
 As discussed in \cite{thiemann1998quantumV,thiemann1998kinematical}, the smeared Hamiltonian constraint in terms of $\theta_A(x)$ is
\begin{equation}
H_F[N]:=\sum_{x\in\Sigma} N(x) H_F(x)
\end{equation}
where $H_F(x)$ is given by         
\begin{equation}
\begin{aligned}
H_F=&i\frac{1}{2\sqrt{q}}(\theta^\dagger E_i^a\sigma^i D_a\theta-(D_a\theta)^\dagger E_i^a\sigma^i\theta)-\beta\frac{1}{2\sqrt{q}} E^a_i K_a^i \theta^\dagger\theta-\frac{1+\beta^2}{\beta}\frac{1}{2\sqrt{q}}D_a E^a_i \theta^\dagger\sigma^i\theta-\beta\frac{1}{2\sqrt{q}} E^a_iD_a\Big( \theta^\dagger\sigma^i\theta\Big).
\end{aligned}
\end{equation}

\begin{figure}
\centering
\includegraphics[width=0.4\textwidth]{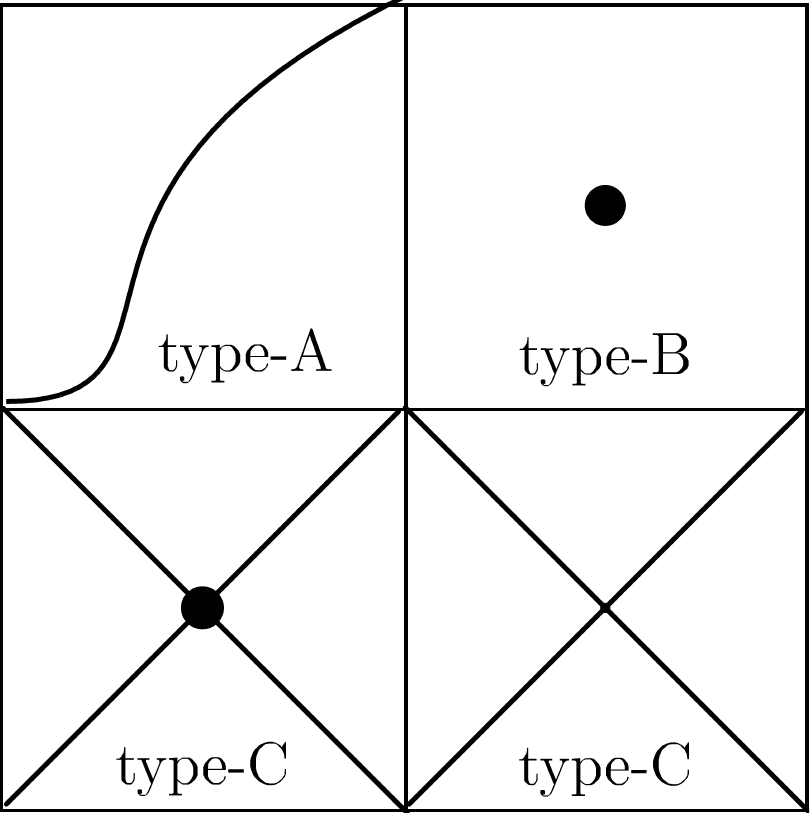}
\caption{Different types of graphs where in the lower left panel is a type-C graph with a fermionic vertex and, the lower right panel, a type-C graph with a gravitational vertex.}\label{fig:types}
\end{figure}

Fix a coordinate system $x^a$ on $\Sigma$ and a positive number $\epsilon$. Divide $\Sigma$ into a family $\mathcal C_\epsilon$ of cells such that each cell $C\in \mathcal C_\epsilon$ is cubic with the coordinate volume less than $\epsilon^3$, and that different cells can only share points on their boundaries. 
Given a graph $\gamma=\gamma_G\cup\gamma_F$, for each cell $C\in \mathcal C_\epsilon$,  let $\gamma_C$ denote $\gamma\cap C$. Since the limit  $\epsilon\to 0$ will be considered eventually,  we will assume that $\epsilon$ is small enough such that $\gamma_C\neq \emptyset$ is one of the following types (see Fig. \ref{fig:types}):
\begin{itemize}
\item[(i)] the type-A graph: $\gamma_C$ is composed of a single edge;
\item[(ii)] the type-B graph: $\gamma_C$  is composed of a single fermionic vertex without  connecting any edges;
\item[(iii)] the type-C graph: $\gamma_C$ is composed of edges intersecting a single vertex. 
\end{itemize} 

For each cell $C\in\mathcal  C_\epsilon$, let us define
\begin{equation}\label{eq:HCi}
\begin{aligned}
H_C^{(1)}=&\int_C\dd^3 xN(x)(D_a\theta)^\dagger(x)E^a_i(x)\sigma^i \theta(x),\\
H_C^{(2)}=&\int_C\dd^3 x N(x)E^a_i(x)K_a^i(x)\theta^\dagger(x)\theta(x),\\
H_C^{(3)}=&\int_C\dd^3 x N(x)D_aE^a_i(x)\theta^\dagger(x)\sigma^i\theta(x),
\end{aligned}
\end{equation}
and introduce
\begin{equation}\label{eq:HCepsilon}
H_{\mathcal C_\epsilon}=\sum_{C\in\mathcal C_\epsilon}\frac{1}{V_C}\left(-iH_C^{(1)}-\frac{\beta}{2} H_C^{(2)}-\frac{1+\beta^2}{2\beta}H_C^{(3)}-\beta H_C^{(1)}\right),
\end{equation}
%
%
where the volume $V_C$ of $C$ is $V_C=\int_C\dd^3 x\sqrt{|\det(E)|}.$
Then $H_F[N]$ is the limit of $\frac{1}{2}H_{\mathcal C_\epsilon}$ plussing its complex conjugate as $\epsilon\to 0$, i.e. 
\begin{equation}\label{eq:limitH}
H_F[N]=\frac{1}{2}\lim_{\epsilon\to 0}(H_{\mathcal C_\epsilon}+c.c.).
\end{equation}

To quantize $H_{\mathcal C_\epsilon}$, we need to promote
\begin{equation}\label{eq:tildeH}
\tilde H_C^{(i)}=\frac{1}{V_C}H_C^{(i)},\ i=1,2,3,\ C\in\mathcal C_\epsilon,
\end{equation} 
 to an operator. As known in LQG, the volume operator is not invertible.  We thus need to regularize $V_C^{-1}$  as in  \cite{yang2016new}  to be $V_C^{-1}=\left(\sqrt{V_C^{-1}}\right)^2$ with
 \begin{equation}
\begin{aligned}
\sqrt{V_C^{-1}}=\kappa_0' \frac{4\times 8}{6}\left(\frac{2}{\kappa\beta}\right)^3\sum_{e,e',e'' \text{ at } v}\epsilon(e,e',e'')\tr\left(h_e^{-1}\{h_e,V_C\}h_{e'}^{-1}\{h_{e'},V_C\}h_{e''}^{-1}\{h_{e''},V_C\}\right).
\end{aligned}
\end{equation}
where $\epsilon(e,e',e'')=0,\pm 1$ depending on the orientation  of $e\wedge e'\wedge e''$, and $\kappa_0'$ is a constant to remove the dependence of partition. 
Then, the operator $\widehat{\tilde H_C^{(i)}}$ is given by
\begin{equation}\label{eq:tildeHCoperator}
\widehat{\tilde H_C^{(i)}}=\widehat{\sqrt{V_C^{-1}}}\hat H_C^{(i)}\widehat{\sqrt{V_C^{-1}}},
\end{equation}
where $\hat H_C^{(i)}$  will be discussed latter, and the  operator $\widehat{\sqrt{V_C^{-1}}}$ reads
\begin{equation}\label{eq:squarerootV}
\begin{aligned}
\widehat{\sqrt{V_C^{-1}}}=\kappa_0'\frac{4\times 8}{6}\left(\frac{2}{i\hbar\kappa\beta}\right)^3\sum_{v \text{ in } C}\sum_{e,e',e'' \text{ at } v}\epsilon(e,e',e'')\times\\
  {\tr}\left(h_e^{-1}[h_e,\hat V_C]h_{e'}^{-1}[h_{e'},\hat V_C]h_{e''}^{-1}[h_{e''},\hat V_C]\right).
\end{aligned}
\end{equation}
Here there are several issues on $\widehat{\sqrt{V_C^{-1}}}$. At first, $\widehat{\sqrt{V_C^{-1}}}$  actually acts on states  $\psi_{\gamma_C}$ with respect to  the graph $\gamma_C=\gamma\cap C$. Given a state $\psi_{\gamma_C}$, according to \eqref{eq:decompsitionfull}, there is a unique graph $\mathbf{P}(\gamma_C)\subset \gamma_C$ such that $\psi_{\gamma_C}\in \mathcal H_{\mathbf{P}(\gamma_C)}^{\rm irr}$. The summation  in \eqref{eq:squarerootV} over $v\in C$ actually means the summation over $v\in V(\mathbf{P}(\gamma_C))$, and the summation over $e,e',e''$ means the summation over $e,e',e''\in E(\mathbf{P}(\gamma_C))$. As a consequence, for $\gamma_C$ being a type-A or type-B graph, one has $\widehat{\sqrt{V_C^{-1}}}\psi_{\gamma_C}=0$, and for $\gamma_C$ being a type-C graph with the vertex $v_C$, one has
\begin{equation}
\begin{aligned}
\widehat{\sqrt{V_C^{-1}}}\psi_{\gamma_C}=\kappa_0'\frac{4\times 8}{6}\left(\frac{2}{i\hbar\kappa\beta}\right)^3\sum_{e,e',e''\in E(\mathbf{P}(\gamma_C))}\epsilon(e,e',e'')\tr\left(h_e^{-1}[h_e,\hat V_{v_C}]h_{e'}^{-1}[h_{e'},\hat V_{v_C}]h_{e''}^{-1}[h_{e''},\hat V_{v_C}]\right)\psi_{\gamma_C},
\end{aligned}
\end{equation}
where $\hat V_{v_C}$ is the volume operator introduced in \cite{ashtekar1997quantumII}. 
 According to this result, the operator $\widehat{\sqrt{V_C^{-1}}}$ at the most right of $\widehat{\tilde H_C^{(i)}}$ will  annihilate the states with respect to the type-A and type-B graphs. Hence, only the states on type-C graphs is needed to be considered. From now on, $\gamma_C$ will be referred to as the type-C graphs until otherwise stated. The vertex for the edges in $\gamma_C$ intersecting will be denoted by $v_C$. 
 
Let us begin with the operator $\widehat{\tilde H_C^{(1)}}$.
Replacing $E_i^a(x)$ by $-i\kappa\hbar\beta\delta/\delta A_a^i(x)$ in $H_C^{(1)}$  in  \eqref{eq:HCi}, we can quantize $H_C^{(1)}$ as
\begin{equation}
\hat H_C^{(1)}=-i\kappa\hbar\beta\int_C\dd^3 x N(x)\widehat{(D_a\theta)}^\dagger(x)\sigma^i\hat\theta(x)\frac{\delta}{\delta A_a^i(x)}.
\end{equation}
Note that we put the derivative $\frac{\delta}{\delta A_a^i(x)}$ at the most right so that the resulting operator is cylindrical consistent, i.e., 
the results of states acted by the resulting operator do not depend on the edges  taking trivial representation.    
Given an edge $e:[0,\delta]\to C$ of $\gamma_C$ with $e(0)=v_C$, let $U_e(t,0,A)$ denote the parallel transport  from $e(0)$ to $e(t)$ along $e$.  Thus, $U_e(t,0,A)$ satisfies 
\begin{equation}\label{eq:holonomy}
\frac{\dd}{\dd t}U_e(t,0;A)=-A_a(e(t))\dot e^a(t)U(t,0;A),\ \text{ and }\ U(0,0;A)=I.
\end{equation}
Defining $h_e:=U(  {\delta,0};A)$, one has, for any $f_a^i(x)$,
\begin{equation}
\begin{aligned}
\int\dd^3 x f_a^i(x) \frac{\delta }{\delta A_a^i(x)} h_e=-\int_0^{  {\delta}}\dd\tau f_a^i(e(\tau))\dot e^a(\tau) U_e(  {\delta},\tau,A)\tau_iU_e(\tau,0,A).
\end{aligned}
\end{equation}
Thus, we get
\begin{equation}\label{eq:operatorinac}
\begin{aligned}
-i\kappa\hbar\beta\int_{C}\dd^3x\,N(x) \widehat{D_a\theta}^\dagger(x) \sigma^i\hat\theta(x)\frac{\delta}{\delta A_a^i(x)} h_e=\kappa\hbar\beta N(s_e)\left(\hat\theta^\dagger(t_e)h_e-\hat\theta^\dagger(s_e)\right)\sigma^i \hat\theta(s_e) \hat J^{v,e}_i h_e
\end{aligned}
\end{equation}
where we used
\begin{equation}\label{eq:psiDpsiE0}
\delta\times \dot e^a(t_1)(D_a\theta)^\dagger (e(t_1))=\theta^\dagger (e(t_1+\delta))U_e(t_1+\delta,t_1;A)-\theta^\dagger (e(t_1))+O(\delta^2).
\end{equation}
Then, for a state $\Psi_C$ with respect to $\gamma_C$, one has
\begin{equation}\label{eq:HC1}
\begin{aligned}
\hat H_C^{(1)}\Psi_C=\kappa\hbar\beta N(v_C) \sum_{e\in\gamma_C}\left(\hat\theta^\dagger(t_e)h_e-\hat\theta^\dagger(v_C)\right)\sigma^i \hat\theta(v_C) \hat J^{v_C,e}_i \Psi_C,
\end{aligned}
\end{equation}
which gives $\widehat{\tilde H^{(1)}_C}$ acting on $\Psi_C$ as
\begin{equation}\label{eq:HC1p}
\begin{aligned}
\widehat{\tilde H^{(1)}_C}\Psi_C=\kappa\hbar\beta  N(v_C)\sum_{e\in\gamma_C}\widehat{\sqrt{V_C^{-1}}}\left(\hat\theta^\dagger(t_e)h_e-\hat\theta^\dagger(v_C)\right)\sigma^i \hat\theta(v_C) \hat J^{v_C,e}_i \widehat{\sqrt{V_C^{-1}}}\Psi_C.
\end{aligned}
\end{equation} 

For the second term $\widehat{\tilde H_C^{(2)}}$, one has 
 \begin{equation}
\hat H_C^{(2)}=\int_C\dd^3 x N(x)E^a_i(x)K_a^i(x)\theta^\dagger(x)\theta(x).
 \end{equation}
 Taking advantage of the Thiemann's trick to quantize  pure-gravity Hamiltonian constraint in LQG \cite{thiemann2007modern}, one has
 \begin{equation}
\int_{C}\dd^3 xf(x) E^a_i(x)K_a^i(x)=\frac{1}{2\kappa\beta^2}\left\{\int_C\dd^3 xf(x)H_E(x),V_C\right\},
 \end{equation} 
 where $H_E(x)$ is the Euclidean part of the pure-gravity Hamiltonian constraint, i.e.
  $$H_E(x)=\frac{\epsilon_{ijk}F_{ab}^i(x)E^a_j(x)E^b_k(x)}{\sqrt{q(x)}}.$$ 
  Thus, $H_C^{(2)}$ is quantized as
 \begin{equation}\label{eq:HC2}
 \begin{aligned}
 \hat H_C^{(2)}=\frac{1}{2i\kappa\hbar\beta^2}\left[N(v_C)\hat{H}_{E,v_C}\hat\theta^\dagger(v_C)\hat\theta(v_C),\hat V_{v_C}\right]
 \end{aligned}
 \end{equation}
 where $\hat{H}_{E,v_C}$ denotes the Euclidean part of the pure-gravity Hamiltonian constraint operator at $v_C$, and $\hat V_{v_C}$, the volume operator at $v_C$.  Eq. \eqref{eq:HC2} leads to the operator $\widehat{\tilde H^{(2)}_C}$ given by
 \begin{equation}\label{eq:HC2p}
 \begin{aligned}
 \widehat{\tilde H^{(2)}_C}=\frac{1}{2i\kappa\hbar\beta^2}}N(v_C)\sqrt{\widehat{V_{v_C}^{-1}}}\left(\hat{H}_{E,v_C}\hat V_{v_C}-\hat V_{v_C}\hat{H}_{E,v_C}\right)\hat\theta^\dagger(v_C)\hat\theta(v_C)\sqrt{\widehat{V_{v_C}^{-1}}.
 \end{aligned}
 \end{equation}
Finally, for the third term $\widehat{\tilde H_C^{(3)}}$, we have
\begin{equation}\label{eq:HC3}
\begin{aligned}
\hat H_C^{(3)}\Psi_C=\kappa\hbar\beta N(v_C)\left(\sum_{e \text{ at } v_C}\hat J_i^{v_C,e}\right)\hat\theta^\dagger(v_C)\sigma^i\hat\theta(v_C)\Psi_C,
\end{aligned}
\end{equation}
by taking advantage of the Gaussian constraint operator of the pure gravity.  Thus $\widehat{\tilde H_C^{(3)}}$ reads
\begin{equation}\label{eq:HC3p}
\begin{aligned}
\widehat{\tilde H_C^{(3)}}\Psi_C=\kappa\hbar\beta N(v_C)\widehat{\sqrt{V_C^{-1}}}\left(\sum_{e \text{ at } v_C}\hat J_i^{v_C,e}\right)\hat\theta^\dagger(v_C)\sigma^i\hat\theta(v_C)\widehat{\sqrt{V_C^{-1}}}\Psi_C
\end{aligned}
\end{equation}
With \eqref{eq:HC1p}, \eqref{eq:HC2p} and \eqref{eq:HC3p}, the operator $\widehat{H_{\mathcal C_\epsilon}}$ is 
\begin{equation}\label{eq:HCall}
\widehat{H_{\mathcal C_\epsilon}}=\sum_{C\in \mathcal C_\epsilon^{(3)}} -i\widehat{\tilde H_C^{(1)}}-\frac{\beta}{2} \widehat{\tilde  H_C^{(2)}}-\frac{1+\beta^2}{2\beta}\widehat{\tilde  H_C^{(3)}}-\beta \widehat{\tilde  H_C^{(1)}}
\end{equation}
where $$\mathcal C_\epsilon^{(3)}=\{C\in \mathcal C_\epsilon, \gamma\cap C \text{ is type-C}\}.$$  

Even though $\widehat{H_{\mathcal C_\epsilon}}$ in \eqref{eq:HCall} is defined with a partition structure $\mathcal C_\epsilon$ on $\Sigma$, this partition structure is indeed not necessary. One can define an operator equivalent to $\widehat{H_{\mathcal C_\epsilon}}$ with $\Sigma$ endowed with  another structure,  which is more convenient for the further study.  To this end, let us first introduce the following notion. 
 \begin{defi}[removable vertex] A vertex $v$ of a graph $\gamma$ is removable if it satisfies the following conditions.
\begin{itemize}
\item[(i)] $v$ is a bivalence vertex connecting $e_1$ and $e_2$;
\item[(ii)] The composition of $e_1$ and $e_2$ as a curve is $C^m$ and semianalytic. 
\end{itemize}
\end{defi}
Given a graph $\gamma$, one can obtain another graph $\ker(\gamma)$ by removing all of its removable vertices. $\ker(\gamma)$ will be called the kernel of $\gamma$. Let $\Gamma_{\ker}$ be the collection of the kernels of all graphs in $\Gamma_o$. Fix once and for all a parametrization for each $\gamma\in\Gamma_{\ker}$, where a parametrization of a graph is an assignment to each edge $e\in E(\gamma)$ a parametrization $[0,1]\ni t\to e(t)\in\Sigma$. 
For an edge $e$ of a kernel in $\Gamma_{\ker}$ taking $v$ as an endpoint, we can define $e(v,\delta)\subset e$ as the segment starting from $v$ and ending at either $e(\delta)$ for $v=e(0)$ or $e(1-\delta)$ for $v=e(1)$.  Given a graph  $\gamma=\gamma_G\cup\gamma_F$, for each $v\in V(\gamma_G)$, mimicking the operator 
$\widehat{\tilde H_C^{(i)}}$ in  \eqref{eq:HC1p}, \eqref{eq:HC2p} and \eqref{eq:HC3p}, we define
\begin{equation}\label{eq:expressionH123}
\widehat{\tilde H^{(i)}(v;\vec\delta_v)}=\widehat{\sqrt{V_v^{-1}}}\widehat{H^{(i)}(v;\vec\delta_v)}\widehat{\sqrt{V_v^{-1}}},\ i=1,2,3
\end{equation}
with
 \begin{equation}\label{eq:expressionH123p}
 \begin{aligned}
 \widehat{H^{(1)}(v;\vec\delta_v)}=&\kappa\hbar\beta  N(v)\sum_{e \text{ at } v}\left(\hat\theta^\dagger(t_{e(v,\delta_{v,e})})h_{e(v,\delta_{v,e})}\sigma^i\hat\theta(v) -\hat\theta^\dagger(v)\sigma^i \hat\theta(v)\right) \hat J^{v,e}_i,\\
 \widehat{H^{(2)}(v;\vec\delta_v)}=&\frac{1}{2i\kappa\hbar\beta^2}N(v)\left(\hat{H}_{E,v}\hat V_{v}-\hat V_{v}\hat{H}_{E,v}\right)\hat\theta^\dagger(v)\hat\theta(v),\\
 \widehat{H^{(3)}(v;\vec\delta_v)}=&\kappa\hbar\beta N(v)\left(\sum_{e \text{ at } v}\hat J_i^{v,e}\right)\hat\theta^\dagger(v)\sigma^i\hat\theta(v).
 \end{aligned}
 \end{equation}
 where $\vec\delta_v=\{\delta_{v,e}\}_{e\text{ at } v}$ is a vector of real numbers for each vertex.  
 Taking advantage of these operators, we define 
 \begin{equation}\label{eq:hdelta}
 \begin{aligned}
 \widehat{H(\delta)}:=&\sum_{v\in V(\gamma_G)}- i\widehat{\tilde H^{(1)}(v;\vec\delta_v)}-\frac{\beta}{2} \widehat{\tilde  H^{(2)}(v;\vec\delta_v)}-\frac{1+\beta^2}{2\beta}\widehat{\tilde  H^{(3)}(v;\vec\delta_v)}-\beta \widehat{\tilde  H^{(1)}(v;\vec\delta_v)}\\
 =&\sum_{v\in V(\gamma)}- i\widehat{\tilde H^{(1)}(v;\vec\delta_v)}-\frac{\beta}{2} \widehat{\tilde  H^{(2)}(v;\vec\delta_v)}-\frac{1+\beta^2}{2\beta}\widehat{\tilde  H^{(3)}(v;\vec\delta_v)}-\beta \widehat{\tilde  H^{(1)}(v;\vec\delta_v)}
 \end{aligned}
 \end{equation}
 where $\delta\equiv \{\vec\delta_v\}_{v\in V(\gamma)}$ and the second step is a consequence of the operator $\sqrt{\widehat{V_v^{-1}}}$ in $\widehat{\tilde H^{(i)}(v;\vec\delta_v)}$. Actually, due to the operator $\sqrt{\widehat{V_v^{-1}}}$ in $\widehat{\tilde H^{(i)}(v;\vec\delta_v)}$, $\widehat{\tilde H^{(i)}(v;\vec\delta_v)}$ with $v\notin\gamma_G$ vanish for all $i=1,2,3$. Comparing the operators  $\widehat{H(\delta)}$  and $\widehat{H_{\mathcal C_\epsilon}}$, one can verify easily that $\widehat{H(\delta)}=\widehat{H_{\mathcal C_\epsilon}}$ with a suitably chosen $\delta$. Hence, we can use $\widehat{H(\delta)}$ instead of $\widehat{H_{\mathcal C_\epsilon}}$ for our further study. In the rest of the paper, we will assume that $\delta_{v,e}$ is a constant for all $v$ and $e$ for convenience. The discussion for non-constant $\delta$ can be discussed similarly.  
  
When the limit of $\widehat{H(\delta)}$ as $\delta\to 0$ is taken, the only nontrivial term in \eqref{eq:expressionH123p} one needs to consider is 
 $$\hat{\mathfrak h}_1(v,e;\delta):=\hat\theta^\dagger(t_{e(v,\delta)})h_{e(v,\delta)}\sigma^i  \hat\theta(v)\hat J_i^{v,e}.$$ 
 By definition, we have
 \begin{equation}\label{eq:interestingH}
 \begin{aligned}
 &\hat{\mathfrak h}_1(v,e;\delta)\left(D^j_{mn}(h_{e(v,\delta)})\otimes |k_1,k_2\rangle_v\right)\\
 =&-2 W_jW_{\frac12}\sum_{J=j\pm \frac12}\sum_{A,C}(-1)^{m+A-n-C}
 \begin{bmatrix}
 \frac12&\frac12&1\\
 j&j&J
 \end{bmatrix}
 \begin{pmatrix}
 j&\frac12&J\\
 m&A&-(m+A)
 \end{pmatrix}
 \begin{pmatrix}
 J&\frac12&j\\
 -(n+C)&C&n
 \end{pmatrix}\times\\
& D^J_{m+A,n+C}(h_{e(v,\delta)})\otimes\hat\theta_C(v)|k_1,k_2\rangle_v\otimes \hat\theta_A^\dagger(t_{e(v,\delta)})|0,0\rangle_{t_{e(v,\delta)}},
 \end{aligned}
 \end{equation}
 where $W_j=\sqrt{j(j+1)(2j+1)}$, $ \begin{bmatrix}
 J&j&\frac12\\
 \frac12&1&J
 \end{bmatrix}$ is the $6j$ symbol and $ \begin{pmatrix}
 j_1&j_2&j_3\\
 m_1&m_2&m_3
 \end{pmatrix}$ denotes the $3j$ symbol. Eq. \eqref{eq:interestingH} gives us the picture that the operator $\hat{\mathfrak h}_1(v,e;\delta)$ moves the fermionic vertex at $v$ to $t_{e(v,\delta)}\in e$, and simultaneously change the spin on $e(v,\delta)$. To illustrate this statement, let us use a disk $ \makeSymbol{\raisebox{0.0\height}{\includegraphics[width=0.015\textwidth]{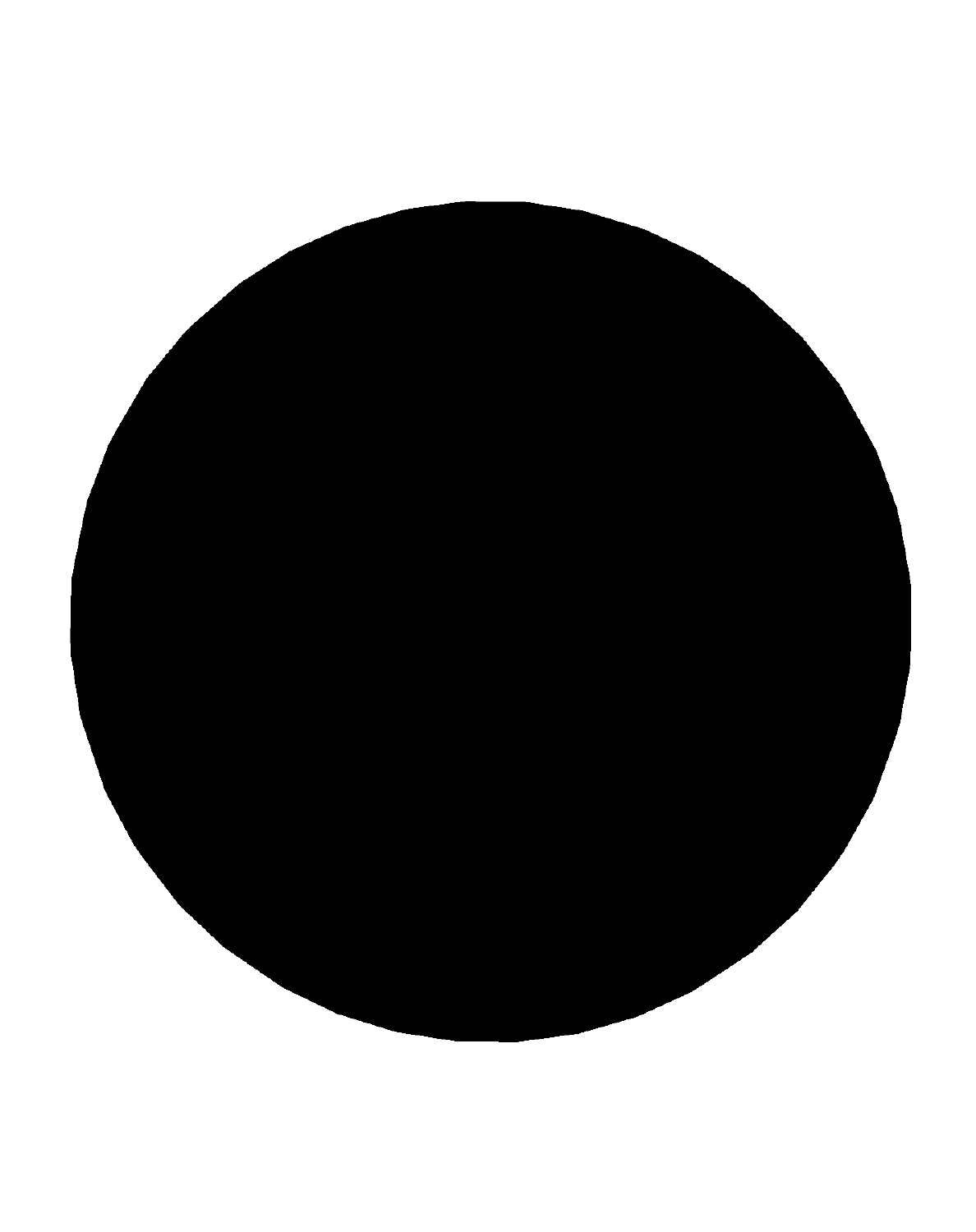}}}$ to represent a fermionic vertex and a solid line to represent the edge $e$. Then, by omitting the explicit coefficients in the right hand of \eqref{eq:interestingH}, one has
 \begin{itemize}
 \item[(1)] for $k_1=1=k_2$, 
 \begin{equation}\label{eq:movingvertex1}
 \begin{aligned}
  &\hat{\mathfrak h}_1(v,e;\delta) \makeSymbol{\raisebox{0.0\height}{\includegraphics[width=0.15\textwidth]{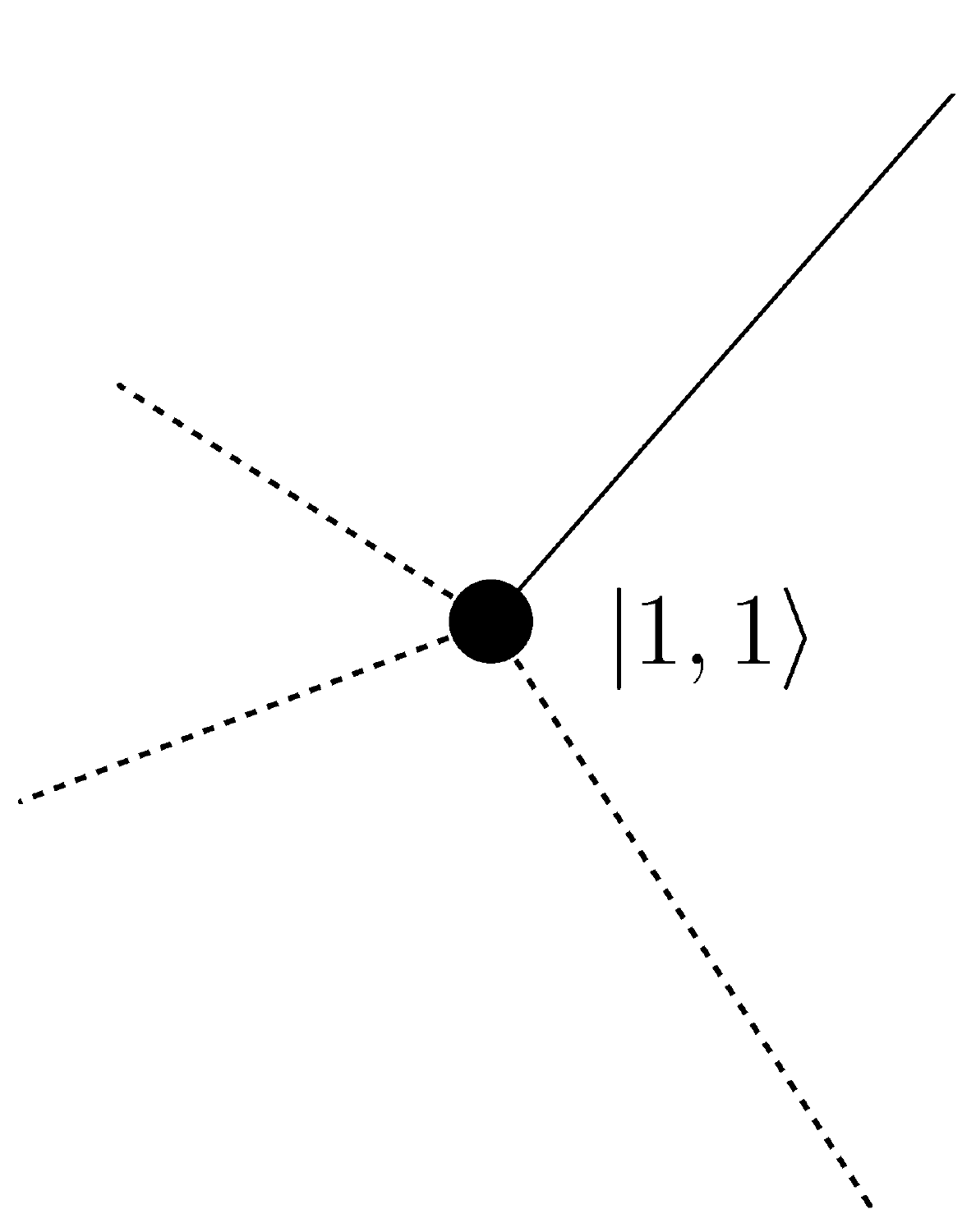}}}\to\sum_{(i_1,i_2)\in\{(1,0),(0,1)\}}\makeSymbol{\raisebox{0.0\height}{\includegraphics[width=0.15\textwidth]{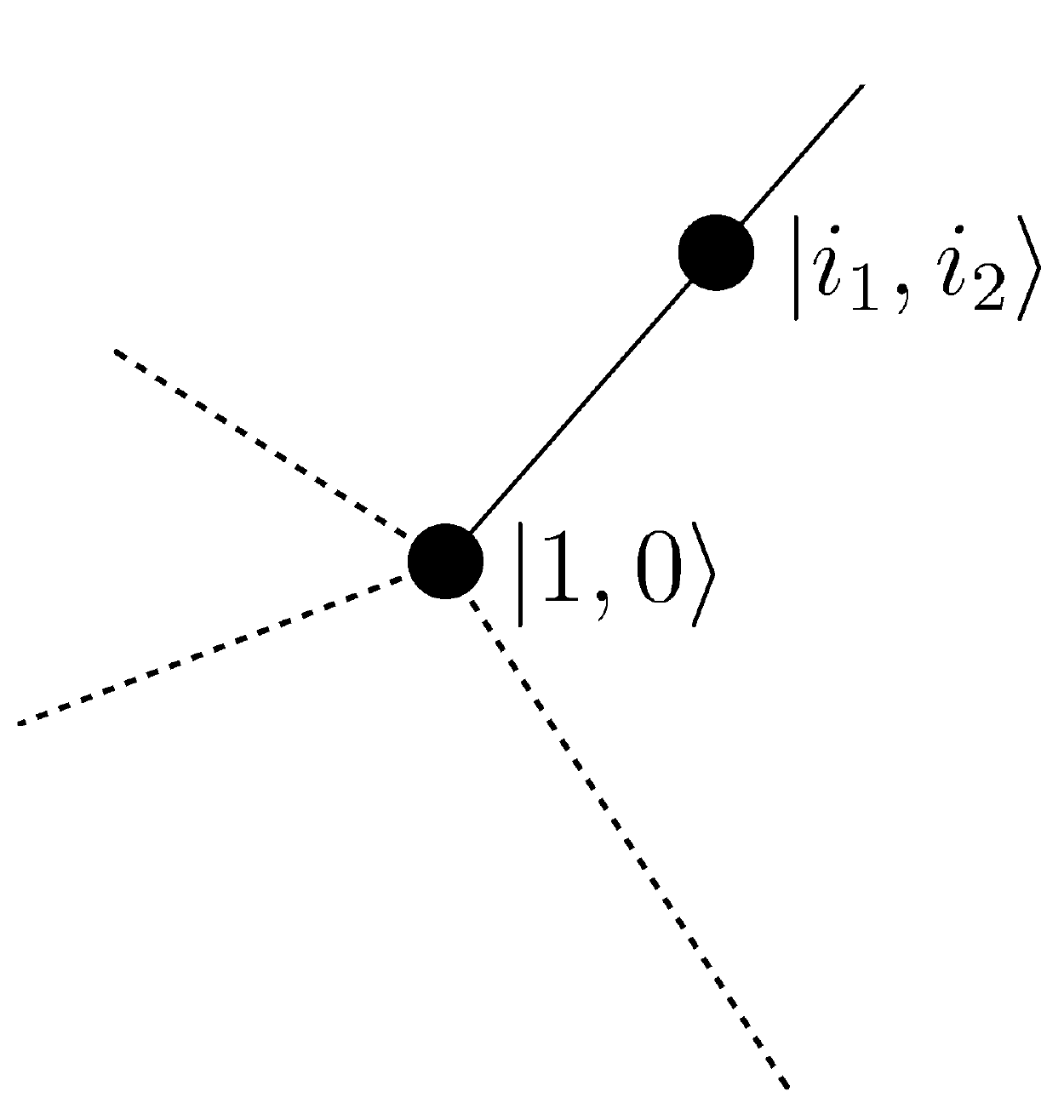}}}+\sum_{(i_1,i_2)\in\{(1,0),(0,1)\}}\makeSymbol{\raisebox{0.0\height}{\includegraphics[width=0.15\textwidth]{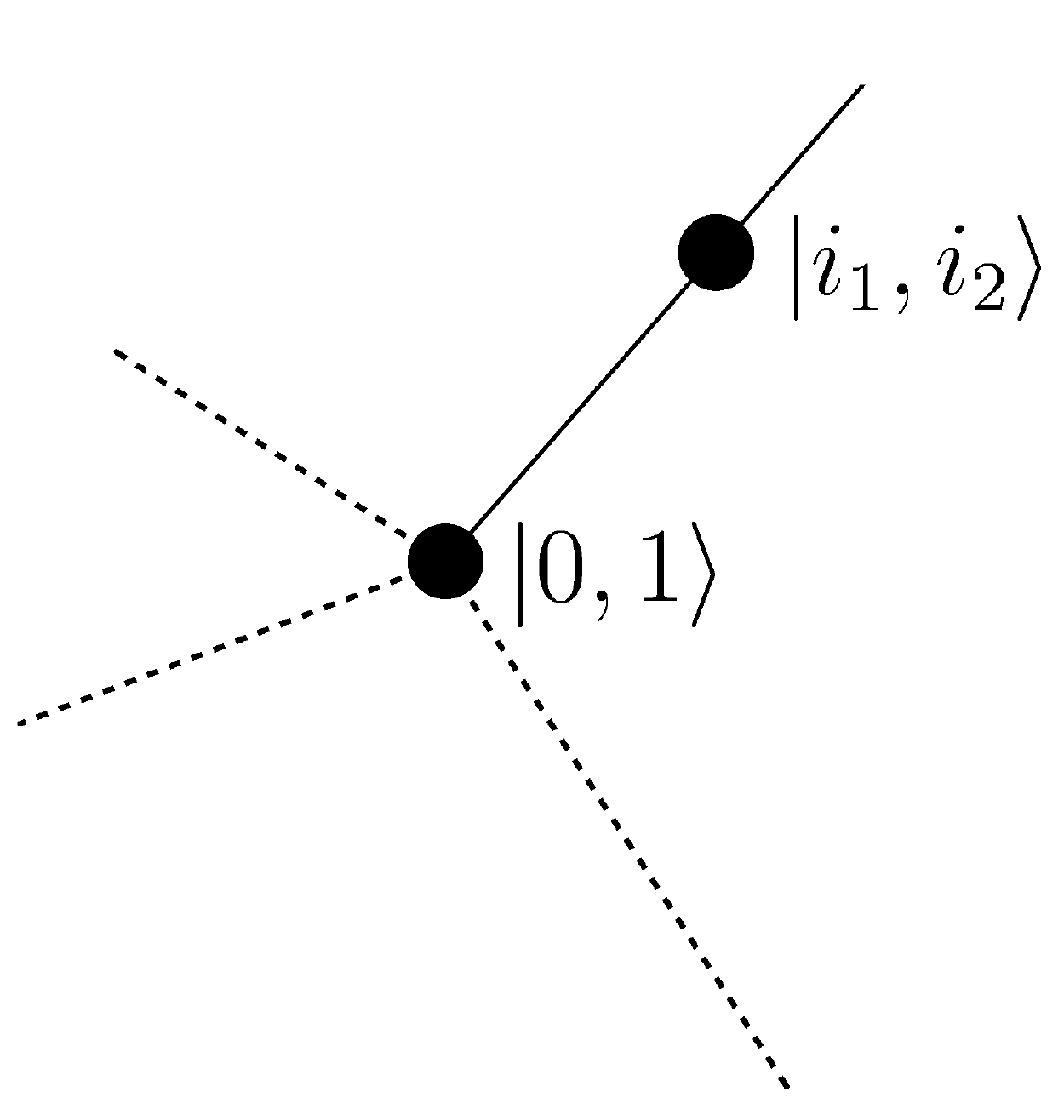}}}\\
 \end{aligned}
 \end{equation}
 \item[(2)] for $k_1=1$ and $k_2=0$
 \begin{equation}\label{eq:movingvertex2}
 \begin{aligned}
 \hat{\mathfrak h}_1(v,e;\delta)\makeSymbol{\raisebox{0.0\height}{\includegraphics[width=0.15\textwidth]{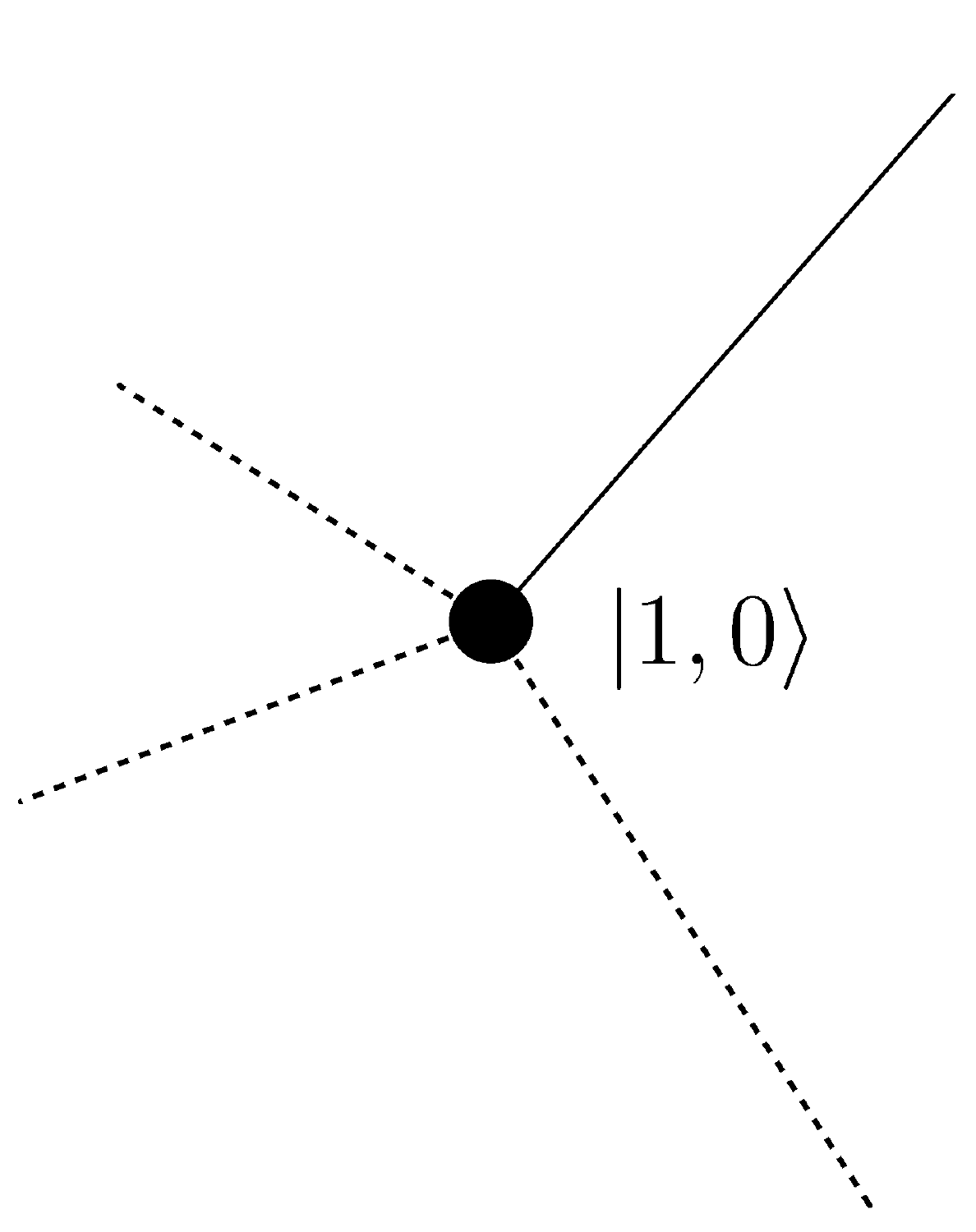}}}\to&\sum_{(i_1,i_2)\in\{(1,0),(0,1)\}}\makeSymbol{\raisebox{0.0\height}{\includegraphics[width=0.15\textwidth]{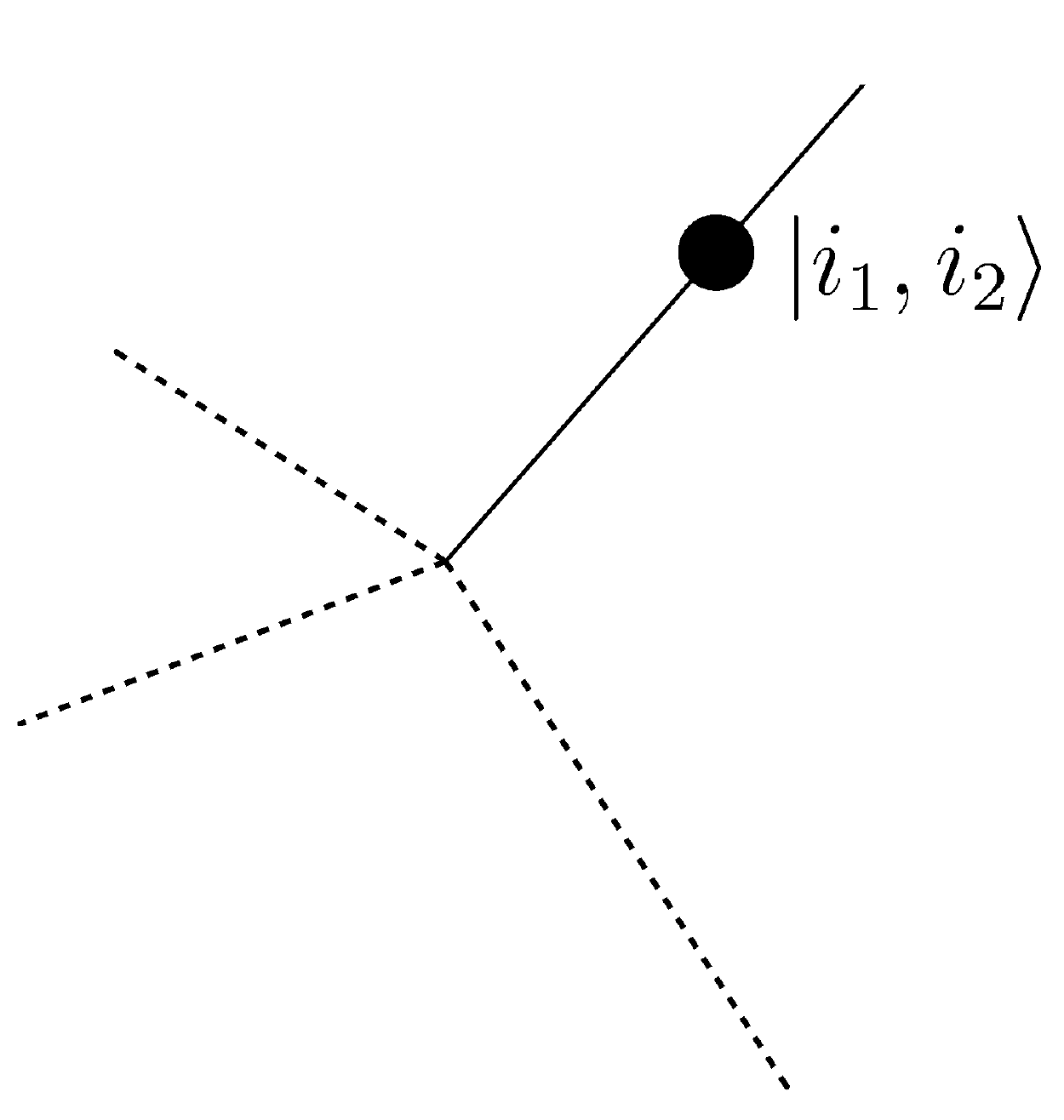}}}
 \end{aligned}
 \end{equation}
 \item[(3)] for $k_1=0$ and  $k_2=1$
 \begin{equation}\label{eq:movingvertex3}
 \begin{aligned}
 \hat{\mathfrak h}_1(v,e;\delta)\makeSymbol{\raisebox{0.0\height}{\includegraphics[width=0.15\textwidth]{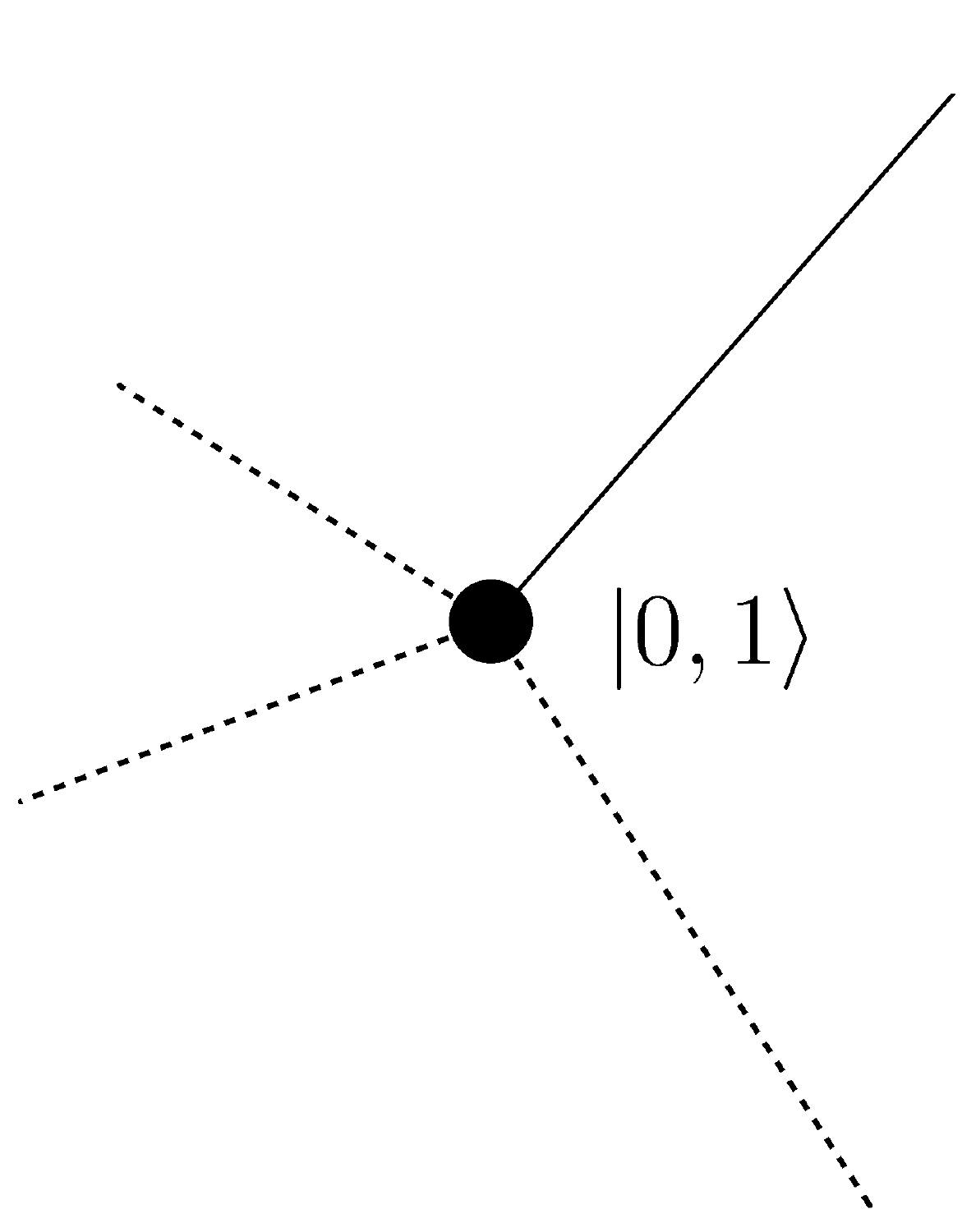}}}\to&\sum_{(i_1,i_2)\in\{(1,0),(0,1)\}}\makeSymbol{\raisebox{0.0\height}{\includegraphics[width=0.15\textwidth]{hamiltonianaction5}}}
 \end{aligned}
 \end{equation}
 \end{itemize}
 According to \eqref{eq:movingvertex1}, \eqref{eq:movingvertex2} and \eqref{eq:movingvertex3}, one gets $\langle \hat{\mathfrak h}_1(v,e;\delta)\psi|\hat{\mathfrak h}_1(v,e;\delta')\psi\rangle=0$ for $\delta\neq\delta'$, which implies that the limit of $\hat{\mathfrak h}_1(v,e;\delta)$ as $\delta\to 0$ does not exist. As a consequence, the limit of $\widehat{H(\delta)}$ as $\delta\to 0$ does not exist, too. However, as final operator should be define as $\lim_{\delta\to 0}\widehat{H(\delta)}$, we need to introduce the  vertex Hilbert space $\mathcal H_{\rm vtx}$ which  is in the dual space of $\cylt$. Once $\mathcal H_{\rm vtx}$ is introduced, $\widehat{H(\delta)}$ can be promoted to an operator $\widehat{H(\delta)}{}^*$ in $\mathcal H_{\rm vtx}$ by the duality such that $\widehat{H(\delta)}{}^*=\widehat{H(\delta')}{}^*$ for $\delta\neq \delta'$. Then, the limit of $\widehat{H(\delta)}{}^*$ as $\delta\to 0$ exists. It will be seen below that  $\mathcal H_{\rm vtx}$ is defined to be the kinematical Hilbert space averaged by diffeomorphisms preserving some particular vertices.    
 
\begin{figure}
 \centering
 \includegraphics[width=0.5\textwidth]{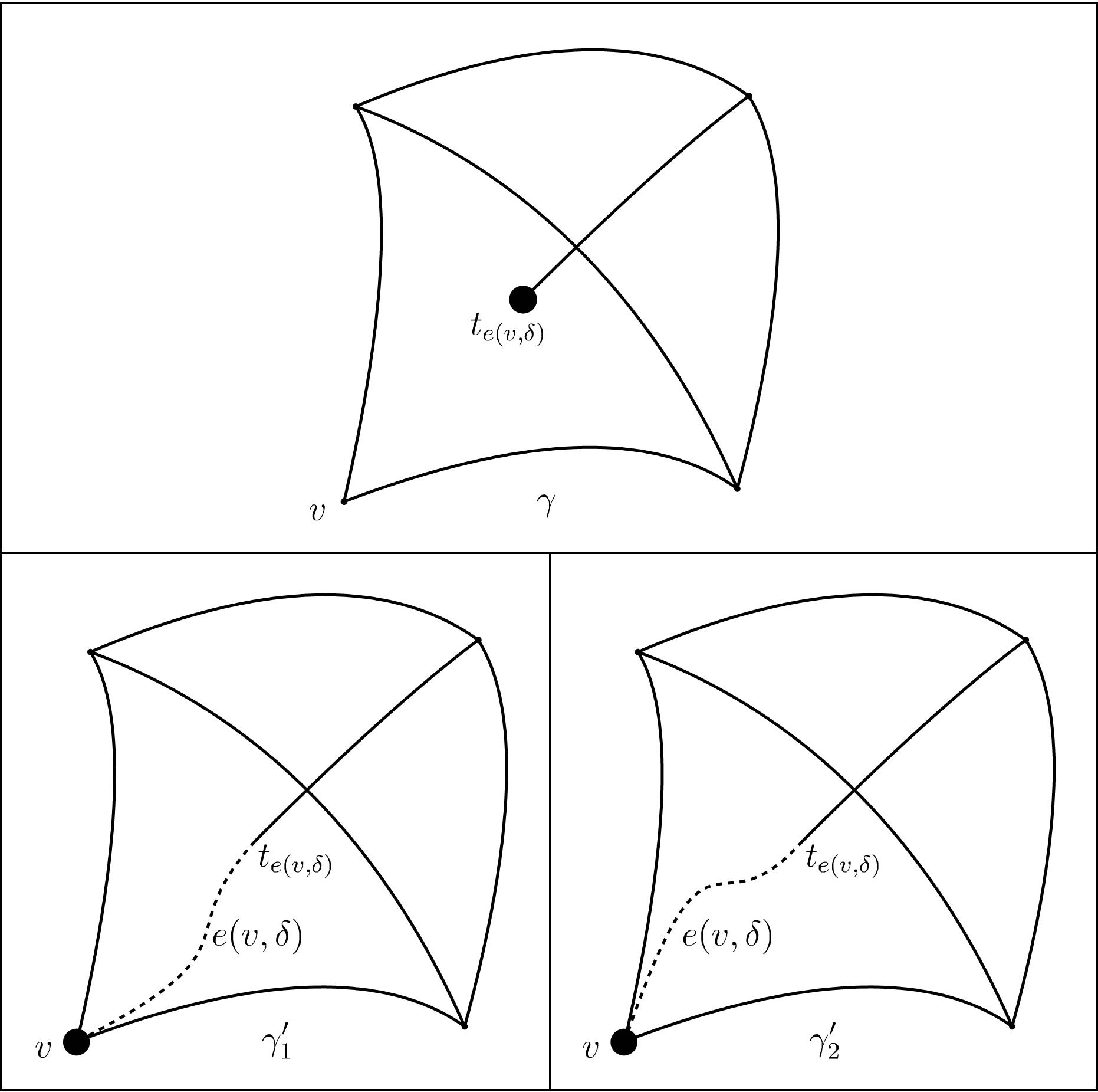}
 \caption{An example of a graph $\gamma$ (top panel) such that $\mathcal H_\gamma^{\rm irr}$ is not in the domain of $\widehat{H^{(1)}(v;\delta)}{}^\dagger$, and examples of graphs $\gamma'$ (bottom panel) such that $\langle \widehat{H^{(1)}(v;\delta)}{}^\dagger\Psi_\gamma|\Phi_{\gamma'}\rangle\neq 0$. The black disks represent fermionic vertices. 
  }\label{fig:adjointg}
 \end{figure}

 Another issue motivating us to introduce the vertex Hilbert space is that the adjoint operator to $\widehat{H(\delta)}$ is not densely defined. To see this, let us investigate the adjoint of $\widehat{H^{(1)}(v;\delta)}$, a term of $\widehat{H(\delta)}$. By definition, for a state $\Psi_\gamma\in \mathcal H_{\gamma}^{\rm irr}$ with $\gamma$ taking $v$ as a vertex, the adjoint $\widehat{H^{(1)}(v;\delta)}{}^\dagger$ acts on it  such that for all $\Phi_{\gamma'}\in \mathcal H$,
  \begin{equation}\label{eq:h1dagger}
  \begin{aligned}
 & \langle \widehat{H^{(1)}(v;\delta)}{}^\dagger\Psi_\gamma|\Phi_{\gamma'}\rangle=\langle \Psi_\gamma| \widehat{H^{(1)}(v;\delta)}\Phi_{\gamma'}\rangle\\
=&\sum_{
\substack{e\in E(\gamma')\\ e\text{ at }v}}\kappa\hbar\beta N(v)\left\langle  (\hat{\mathfrak h}_1(v,e;\delta)^\dagger-\hat\theta(v)\sigma^i\hat\theta(v)\hat J_i^{v,e}) \Psi_\gamma\middle|\Phi_{\gamma'}\right\rangle
  \end{aligned}
 \end{equation}
 with $\hat{\mathfrak h}_1(v,e;\delta)^\dagger=\hat J_i^{v,e} \hat\theta^\dagger(v)\sigma^i h_{e(v,\delta)}^{-1} \hat\theta(t_{e(v,\delta)})$. The subtlety arises due to the summation over edges of $\gamma'$ rather than $\gamma$. Let us consider a state $\Psi_\gamma$ with $\gamma$ shown in Fig. \ref{fig:adjointg}. Moreover,  the fermion state of $\Psi_\gamma$ at the fermionic vertex $t_{e(v,\delta)}$ is chosen to be $a|0,1\rangle_{t_{e(v,\delta)}}+b|1,0\rangle_{t_{e(v,\delta)}}$ for some $a,b\in\mathbb C$. Then, for each graph $\gamma'$ obtained from $\gamma$ by using an edge $e(v,\delta)$ to connect the vertex $t_{e(v,\delta)}$ and $v$, dropping the fermionic vertex $t_{e(v,\delta)}$ and adding a fermionic vertex at $v$ (see the two examples in Fig.  \ref{fig:adjointg}), one can find a state  $\Phi_{\gamma'}\in \mathcal H_{\gamma'}^{\rm irr}$ such that $\left\langle  (\hat{\mathfrak h}_1(v,e;\delta)^\dagger-\hat\theta(v)\sigma^i\hat\theta(v)\hat J_i^{v,e}) \Psi_\gamma\middle|\Phi_{\gamma'}\right\rangle\neq 0$, where $e\in\gamma'$ is the edge containing $e(v,\delta)$ as a segment. Because there exist uncountably infinitely many such graphs $\gamma'$ and the states $\Phi_{\gamma'}$ associated to different graphs are orthogonal to each other, $\Psi_\gamma$ is not in the domain of $\widehat{H^{(1)}(v;\delta)}{}^\dagger$. 
 Similar argument can be done for all of states in $\mathcal H_{\gamma}^{\rm irr}$ (including  those taking $|1,1\rangle_{t_{e(v,\delta)}}$ at $t_{e(v,\delta)}$), which implies that the entire Hilbert space $\mathcal H_{\gamma}^{\rm irr}$ is not contained in the domain of $\widehat{H^{(1)}(v;\delta)}{}^\dagger$. Besides, it is manifest that there are infinitely many such Hilbert spaces that are excluded from the domain of $\widehat{H^{(1)}(v;\delta)}{}^\dagger$. Actually, according to \eqref{eq:interestingH}, the operator $\hat{\mathfrak{h}}_1(v,e;\delta)$ will change the state taking spin $j=\frac12$ on $e(v,\delta)$ to a superposition of states taking spins $j=0$ and $1$  on $e(v,\delta)$. In other words, the operator $\hat{\mathfrak{h}}_1(v,e;\delta)$  can erase the segment $e(v,\delta)\subset e$, which is indeed the essential  reason for the problem of defining $\widehat{H^{(1)}(v;\delta)}{}^\dagger$. It will be seen from blow that this problem can also be fixed by introducing the vertex Hilbert space to define the limit. After taking the limit as $\delta\to 0$ in the vertex Hilbert space, we will get a projection $\chi_+(|\hat J^{v,e}|^2)$ (see \eqref{eq:corresponding H1}). This projection is left multiplied to the operator  corresponding to $  {\hat{\mathfrak h}_1(v,e;\delta)}$ in $\mathcal H_{\rm vtx}$, and kill the state with $j=0$ on $e(v,\delta)$. 
 
Finally, let us discuss why we choose $H_{\mathcal C_\epsilon}$ rather than its complex conjugate for quantization. Suppose that one choose the complex conjugate of $H_{\mathcal C_\epsilon}$ to do the above quantization. Then, in the resulting operator, one will get a term involving $\hat{\mathfrak h}_1(v,e;\delta)^\dagger=\hat J_i^{v,e} \hat\theta^\dagger(v)\sigma^i h_{e(v,\delta)}^{-1} \hat\theta(t_{e(v,\delta)})$. In contrast to $\hat{\mathfrak h}_1(v,e;\delta)$, $\hat{\mathfrak h}_1(v,e;\delta)^\dagger$ moves the fermionic vertex at $t_{e(v,\delta)}\in e$ to $v$. Now let us consider the states $\hat{\mathfrak h}_1(v,e;\delta)|\psi\rangle$ and $\hat{\mathfrak h}_1(v,e;\delta')|\psi\rangle$ with $\delta\neq\delta'$ for $|\psi\rangle$, says, being $D^{j}_{mn}(h_e)\otimes|k_1,k_2\rangle_v$. Then, one gets  $\hat{\mathfrak h}_1(v,e;\delta)^\dagger(\hat{\mathfrak h}_1(v,e;\delta')|\psi\rangle)=0$ but $\hat{\mathfrak h}_1(v,e;\delta)^\dagger(\hat{\mathfrak h}_1(v,e;\delta)|\psi\rangle)\neq 0$. This result means that $\hat{\mathfrak h}_1(v,e;\delta)^\dagger$ is not diffeomorphism covariant. Namely, diffeomorphism equivalent states, i.e., the states $\hat{\mathfrak h}_1(v,e;\delta)|\psi\rangle$ and $\hat{\mathfrak h}_1(v,e;\delta')|\psi\rangle$,  could be no longer diffeomorphism equivalent after acted by $\hat{\mathfrak h}_1(v,e;\delta)^\dagger$. This facts leads to a problem that $\sum_{e \text{ at }v}\hat{\mathfrak h}_1(v,e;\delta)^\dagger$ cannot be promoted to a well-defined operator in the diffeomorphism invariant Hilbert space.

\subsubsection{the vertex  Hilbert space $\mathcal H_{\rm vtx}$}
Given a graph $\gamma$, we will consider the subgroup of $C^m$ semianalytic diffeomorphisms which act on $V(\ker(\gamma))$ trivially, i.e. the diffeomorphisms preserving every vertex  of $\ker(\gamma)$. This subgroup will be denoted by $\diff_{V(\ker(\gamma))}$. There are two  subgroups of  $\diff_{V(\ker(\gamma))}$. The first one, denoted by $\diff_{\gamma}$, preserves $\gamma$. The other  one, denoted by $\diff^0_{\gamma}$, preserves every edge of $\gamma$. Hence $\diff^0_{\gamma}$ is a subgroup of $\diff_{\gamma}$. 
The quotient 
\begin{equation}
\gs_{\gamma}=\diff_{\gamma}/\diff^0_{\gamma}
\end{equation} 
is the group of graph symmetries of $\gamma$. $\gs_{\gamma}$ is  a finite group with order   {$|\gs_\gamma|$}. Given a state $\Psi_\gamma$ with respect to $\gamma$ and a diffeomorphism $\phi$, $\phi\star\Psi_\gamma$ denotes the pullback of $\Psi$ under $\phi$. The averaging with respect to   {$\gs_\gamma$} defines a projection $P_{\gamma}:\mathcal H_\gamma^{\rm irr}\to \mathcal H_\gamma^{\rm irr}$ , 
\begin{equation}
P_\gamma:\Psi_\gamma\mapsto \frac{1}{|\gs_\gamma|}\sum_{\phi\in\gs_{\gamma}}\phi\star\Psi_\gamma.
\end{equation}
The averaging with respect to the remaining diffeomorphisms $\diff_{V(\ker(\gamma))}/\diff_\gamma$ defines 
\begin{equation}\label{eq:defineeta}
\eta(\Psi_\gamma):=\sum_{\phi\in \diff_{V(\ker(\gamma))}/\diff_\gamma }\phi\star P_\gamma(\Psi_\gamma)=\frac{1}{|\gs_\gamma|}\sum_{\phi\in \diff_{V(\ker(\gamma))}/\diff_\gamma^0 }\phi\star \Psi_\gamma.
\end{equation}
Obviously, $\eta(\Psi_\gamma)$ belongs to the  algebraic dual space $\cylt^*$ of $\cylt$. Taking advantage the decomposition \eqref{eq:decompsitionfull}, one can extend $\eta$ to a well-defined operation on $\cylt$.  Let $\eta(\cylt)\subset \cylt^*$ denote  the image of $\cylt$ under  $\eta$. The group averaging procedure naturally endows  $\eta(\cylt)$ with an inner product
\begin{equation}\label{eq:innerdual}
\left(\eta(\Psi_{\gamma}),\eta(\Psi_{\gamma'}')\right)=(\eta(\Psi_\gamma)|\Psi_{\gamma'}'\rangle,
\end{equation}
where  $(\eta(\Psi_\gamma)|\Psi_{\gamma'}'\rangle$,  the action of $\eta(\Psi_\gamma)\in\cylt^*$ on  $\Psi_{\gamma'}$, is given by
\begin{equation}
(\eta(\Psi_\gamma)|\Psi_{\gamma'}'\rangle=\sum_{\phi\in \diff_{V(\ker(\gamma))}/\diff_\gamma^0 }\langle\phi\star \Psi_{\gamma},\Psi_{\gamma'}'\rangle.
\end{equation}
Here, $\langle\cdot,\cdot\rangle$ is the inner product in $\mathcal H$. Then the vertex Hilbert space $\mathcal H_{\rm vtx}$ is the completion of $\eta(\cylt)$, i.e,
\begin{equation}
\mathcal H_{\rm vtx}=\overline{\eta(\cylt)}.
\end{equation}

Given a finite subset $W\subset \Sigma$, let $\Gamma_{\ker}(W)$ be the collection of graphs $\gamma$ satisfying $V(\ker(\gamma))=W$. Two graphs $\gamma,\gamma'\in \Gamma_{\ker}(W)$  are said to be equivalent, denoted by  $\gamma\sim_d\gamma'$, if there exists a diffeomorphism $\phi\in \diff_W$, i.e., a diffeomorphism preserving $W$, such that $\phi(\gamma)=\gamma'$. Let $[\Gamma_{\ker}(W)]$ be the quotient space $\Gamma_{\ker}(W)/\sim_d$, $[\gamma]\in [\Gamma_{\ker}(W)]$ be the equivalence class of $\gamma$, and $\mathcal S_\gamma^{\rm irr}$ be the image of $P_\gamma$.
Then $\eta:\mathcal S_\gamma^{\rm irr}\to   {\mathcal H}_{\rm vtx}$ is isometric. By $\eta(\mathcal S_\gamma^{\rm irr})$ denoting the image of $\mathcal S_\gamma^{\rm irr}$ under $\eta$, one has
\begin{equation}\label{eq:etaequal}
\eta(\mathcal S_\gamma^{\rm irr})=\eta(\mathcal S_{\gamma'}^{\rm irr}),\ \forall \gamma,\gamma'\in[\gamma]. 
\end{equation}
Thus, we can define 
\begin{equation}
\eta(\mathcal S_{[\gamma]}^{\rm irr}):=\eta(\mathcal S_{\gamma'}^{\rm irr})
\end{equation}
by choosing arbitrary $\gamma'\in [\gamma]$.  
Let ${\rm FS}(\Sigma)$ be the set of finite subsets of $\Sigma$. Then we have
\begin{equation}
\eta(\cylt)=\bigoplus_{W\in {\rm FS }(\Sigma)}\bigoplus_{[\gamma]\in  {[ \Gamma_{\ker}(W)]}}\eta(  {\mathcal S}_{[\gamma]}^{\rm irr})\oplus \mathbb C.
\end{equation}
The factor $1/|\gs_\gamma|$ in \eqref{eq:innerdual} ensures that $\eta$ is an isometric between $\mathcal H_{\gamma}^{\rm irr}$ and $\eta(\mathcal H_\gamma^{\rm irr})$, i.e.,
\begin{equation}\label{eq:isometric}
\eta(\cylt) \cong  \bigoplus_{W\in {\rm FS }(\Sigma)}\bigoplus_{[\gamma]\in   {[ \Gamma_{\ker}(W)]}}  {\mathcal S}_{\sigma([\gamma])}^{\rm irr}\oplus \mathbb C.
\end{equation}
where $\sigma([\gamma])\in[\gamma]$ is a representative of $[\gamma]$ fixed once and for all. 

\subsubsection{the Hamiltonian operator on $\mathcal H_{\rm vtx}$}
Given $\Psi_\gamma\in \mathcal S_\gamma^{\rm irr}$, let $\widehat{H(\delta)}{}^*$ be the dual of $\widehat{H(\delta)}$ acting on $\eta(\Psi_\gamma)$, i.e.,
\begin{equation}
\left(\widehat{H(\delta)}{}^*\eta(\Psi_\gamma)\middle|\Phi_{\gamma'}\right\rangle=\left(\eta(\Psi_\gamma)\middle|\widehat{H(\delta)}\Phi_{\gamma'}\right\rangle, \forall \gamma'\in \Gamma_o,\ \Phi_{\gamma'}\in \mathcal H_{\gamma'}^{\rm irr}. 
\end{equation}
The operator $\widehat{H(\delta)}$ takes the form $\widehat{H(\delta)}=\sum_{v} N(v)  \widehat{\sqrt{V_v^{-1}}}\hat O(v)\widehat{\sqrt{V_v^{-1}}}$ with $\hat O(v)$ representing  some operator. Due to the inverse volume operators, we have
\begin{equation}\label{eq:hstar0}
\begin{aligned}
\lim_{\delta\to 0}\left(\widehat{H(\delta)}{}^*\eta(\Psi_\gamma)\middle|\Phi_{\gamma'}\right\rangle=\lim_{\delta\to 0}\sum_{v\in V(\ker(\gamma))}N(v)\left(\eta(\Psi_\gamma)\middle|\widehat{\sqrt{V_v^{-1}}}\hat O(v)\widehat{\sqrt{V_v^{-1}}}\Phi_{\gamma'}\right\rangle.
\end{aligned}
\end{equation}
Recalling \eqref{eq:hdelta}, we will consider the right hand term by term. 

Let us begin with the operator $\widehat{H^{(1)}(v;\delta)}$ with $v\in V(\ker(\gamma))$. By \eqref{eq:hstar0}, we need to investigate 
\begin{equation}\label{eq:Iv}
I_v(\gamma',\delta)=\left(\eta(\Psi_\gamma)\middle|\widehat{H^{(1)}(v;\delta)}\Phi_{\gamma'}\right\rangle
\end{equation}
for all graphs $\gamma'$ taking $v$ as a vertex. By definition, we can split $\widehat{H^{(1)}(v;\delta)}$ into two parts such that 
\begin{equation}\label{eq:H1dagger}
\widehat{H^{(1)}(v;\delta)}\Phi_{\gamma'}=\kappa \hbar \beta N(v) \hat{\mathfrak{h}}_1(v;\delta)\Phi_{\gamma'}-\kappa \hbar \beta N(v) \hat{\mathfrak{h}}_2(v)\Phi_{\gamma'}
\end{equation}
with $
\hat{\mathfrak{h}}_1(v;\delta)\Phi_{\gamma'}=\sum_{
e\text{ at } v}\hat{\mathfrak h}_1(v,e;\delta)\Phi_{\gamma'}$ and $
\hat{\mathfrak{h}}_2(v)=\sum_{[e] \text{ at } v}\hat\theta^\dagger(v)\sigma^i\hat\theta(v)\hat J_i^{v,[e]}$. Substituting \eqref{eq:H1dagger} into \eqref{eq:Iv}, we will be concerned about  the term
\begin{equation}\label{eq:Iv1}
I_v^{(1)}(\gamma',\delta)=\left(\eta(\Psi_\gamma)\middle|\hat{\mathfrak{h}}_1(v;\delta)\Phi_{\gamma'}\right\rangle=\sum_{\substack{e\in E(\gamma')\\
e\text{ at } v}}\left(\eta(\Psi_\gamma)\middle|\hat{\mathfrak h}_1(v,e;\delta)\Phi_{\gamma'}\right\rangle.
\end{equation}
Substituting \eqref{eq:defineeta} into \eqref{eq:Iv1}, we get
\begin{equation}\label{eq:I1hdagger}
\begin{aligned}
I_v^{(1)}(\gamma',\delta)
=&\sum_{\substack{e\in E(\gamma')\\e\text{ at } v}}\sum_{\phi\in\diff' }\left\langle\hat{\mathfrak h}_1(v,e;\delta)^\dagger(\phi\star\Psi_\gamma)\middle|\Phi_{\gamma'}\right\rangle
\end{aligned}
\end{equation}
with $\diff':=\diff_{V(\ker(\gamma))}/\diff_\gamma$ for abbreviation,
where we employ $\hat{\mathfrak h}_1(v,e;\delta)^\dagger=\hat J_i^{v,e} \hat\theta^\dagger(v)\sigma^i h_{e(v,\delta)}^{-1} \hat\theta(t_{e(v,\delta)})$.  Note that  the summation in the right hand side of \eqref{eq:I1hdagger} is over edges of $\gamma'$, while the operator $\hat{\mathfrak h}_1(v,e;\delta)^\dagger$ acts on states associated to graphs diffeomorphism to $\gamma$. Thus, we \textit{cannot} get $I^{(1)}_v(\gamma',\delta)=\left(\hat{\mathfrak h}_1(v;\delta)^\dagger\eta(\Psi_\gamma)|\Phi_{\gamma'}\right\rangle$ naively from \eqref{eq:I1hdagger}. 

Consider the action of $\hat{\mathfrak h}_1(v,e;\delta)^\dagger$ on $D^{j_e}_{m_en_e}(h_{e(v,\delta)})\otimes D^{j_e'}_{m_e'n_e'}(h_{e\setminus e(v,\delta)})\otimes|k_1,k_2\rangle_{t_{e(v,\delta)}}$, a general state associated to the graph $\{e\}\cup \{t_{e(v,\delta)}\}$, where $e\setminus e(v,\delta)$ is the other segment of $e$ complementing $e(v,\delta)$.  
Here and in the rest, by a state associated to a graph $\alpha$, we refer to a state in $\mathcal H_\alpha^{\rm irr}$. Acting on the state, the holonomy $h_{e(v,\delta)}^{-1}$ in $\hat{\mathfrak h}_1(v,e;\delta)^\dagger$ changes the spin $j_e$ on $e(v,\delta)$ to $j_e\pm \frac12$. Then, for $j_e=\frac12$, the spin on $e(v,\delta)$ is changed to $0$, i.e., the segment $e(v,\delta)$ is erased. Thus, for $j_e=\frac{1}{2}$, acted by the holonomy, the state becomes a combination of one state associated $\{e\}$  and the other state associated to $\{e\setminus e(v,\delta)\}$. However, the operator $\hat J_i^{v,e}$ in  $\hat{\mathfrak h}_1(v;\delta)^\dagger$ will annihilate this one associated to $\{e\setminus e(v,\delta)\}$. Consequently, acted by $\hat{\mathfrak h}_1(v;\delta)^\dagger$, the resulting state is still a state associated to $\{e\}$, regardless of the change of the fermionic vertices by $\hat\theta^\dagger(v)$ and $\hat\theta(t_{e(v,\delta)})$.  

$\phi_o\star\Psi_\gamma$ for some diffeomorphism $\phi_o$ is a state associated to the graph $\phi_o(\gamma)$. Let $\tilde\gamma_G$ and $\tilde\gamma_F$ denote the gravitational sector and the fermionic sector of $\phi_o(\gamma)$, i.e., $\phi_o(\gamma)=\tilde\gamma_G\cup\tilde\gamma_F$. Then, the discussion in the last paragraph tells that $\hat{\mathfrak h}_1(v;\delta)^\dagger\phi_o\star\Psi_\gamma$ is a state associated to some graph whose gravitational sector is $\tilde \gamma_G\cup e(v,\delta)$. Thus, $\left\langle\hat{\mathfrak h}_1(v,e;\delta)^\dagger(\phi_o\star\Psi_\gamma)\middle|\Phi_{\gamma'}\right\rangle\neq 0$ for $\Phi_{\gamma'}\in\mathcal H_{\gamma'}^{\rm irr}$ implies $e\subset \tilde \gamma_G\cup e(v,\delta)$, which means either $e\subset \phi_o(\gamma)$, or $e(v,\delta)\not\subset\phi_o(\gamma)$ but $e\setminus e(v,\delta)\subset \phi_o(\gamma)$. 

Let us first consider the case where $e(v,\delta)\not\subset \phi_o(\gamma)$ but $e\setminus e(v,\delta)\subset \phi_o(\gamma)$ such that $\left\langle\hat{\mathfrak h}_1(v,e;\delta)^\dagger(\phi_o\star\Psi_\gamma)\middle|\Phi_{\gamma'}\right\rangle\neq 0$. The hypothesis $\left\langle\hat{\mathfrak h}_1(v,e;\delta)^\dagger(\phi_o\star\Psi_\gamma)\middle|\Phi_{\gamma'}\right\rangle\neq 0$ implies $\tilde \gamma_G\cup e(v,\delta)=\gamma'_G$ where $\gamma'_G$ is the gravitational sector of $\gamma'$. Together with $e(v,\delta)\not\subset\phi_o(\gamma)$, one gets $\tilde\gamma_G=\gamma'_G\setminus e(v,\delta)$. Since $e(v,\delta)$ is a segment of an edge $e\subset\gamma'_G$, $t_{e(v,\delta)}$ is a 1-valence vertex in $\tilde\gamma_G$. Furthermore, $t_{e(v,\delta)}$ is a 1-valence fermionic vertex because of $\hat{\mathfrak h}_1(v,e;\delta)^\dagger(\phi_o\star\Psi_\gamma)\neq 0$. Thus, $t_{e(v,\delta)}$ is \textit{not} a removable vertex, i.e., all of the diffeomorphisms in $\diff'$ preserve this vertex.  Now, let us come to $\delta'<\delta$. If $\sum_{\phi\in\diff' }\left\langle\hat{\mathfrak h}_1(v,e;\delta')^\dagger(\phi\star\Psi_\gamma)\middle|\Phi_{\gamma'}\right\rangle=\sum_{\phi\in\diff' }\left\langle\hat{\mathfrak h}_1(v,e;\delta')^\dagger(\phi\star\phi_o\star\Psi_\gamma)\middle|\Phi_{\gamma'}\right\rangle$ still does not vanish, there have to exist another diffeomorphism $\phi_o'$ which can move a fermionic vertex in $\phi_o(\gamma)$ to $t_{e(v,\delta')}$ such that $\hat{\mathfrak h}_1(v,e;\delta')^\dagger(\phi_o'\star\phi_o\star\Psi_\gamma)\neq 0$, and simultaneously preserve all of the other vertices but not necessarily $t_{e(v,\delta)}$ in $\phi_o(\gamma)$ such that $\left\langle\hat{\mathfrak h}_1(v,e;\delta')^\dagger(\phi_o'\star\phi_o\star\Psi_\gamma)\middle|\Phi_{\gamma'}\right\rangle\neq 0$. This can be done only if $\phi_o'$ can move $t_{e(v,\delta)}$ to $t_{e(v,\delta')}$ which is impossible because $t_{e(v,\delta)}$ is preserved by all diffeomorphisms in $\diff'$. Therefore, $\sum_{\phi\in\diff' }\left\langle\hat{\mathfrak h}_1(v,e;\delta')^\dagger(\phi\star\Psi_\gamma)\middle|\Phi_{\gamma'}\right\rangle=0$ for all $\delta'<\delta$. This discussion tells that for sufficiently small $\delta$, the case with $e(v,\delta)\not\subset \phi_o(\gamma)$ but $e\setminus e(v,\delta)\subset \phi_o(\gamma)$ can be excluded. Because we finally need to consider the limit as $\delta\to 0$, choosing a sufficiently small $\delta$ can be done without loss of generality.  

\begin{figure}
\centering
\includegraphics[width=0.4\textwidth]{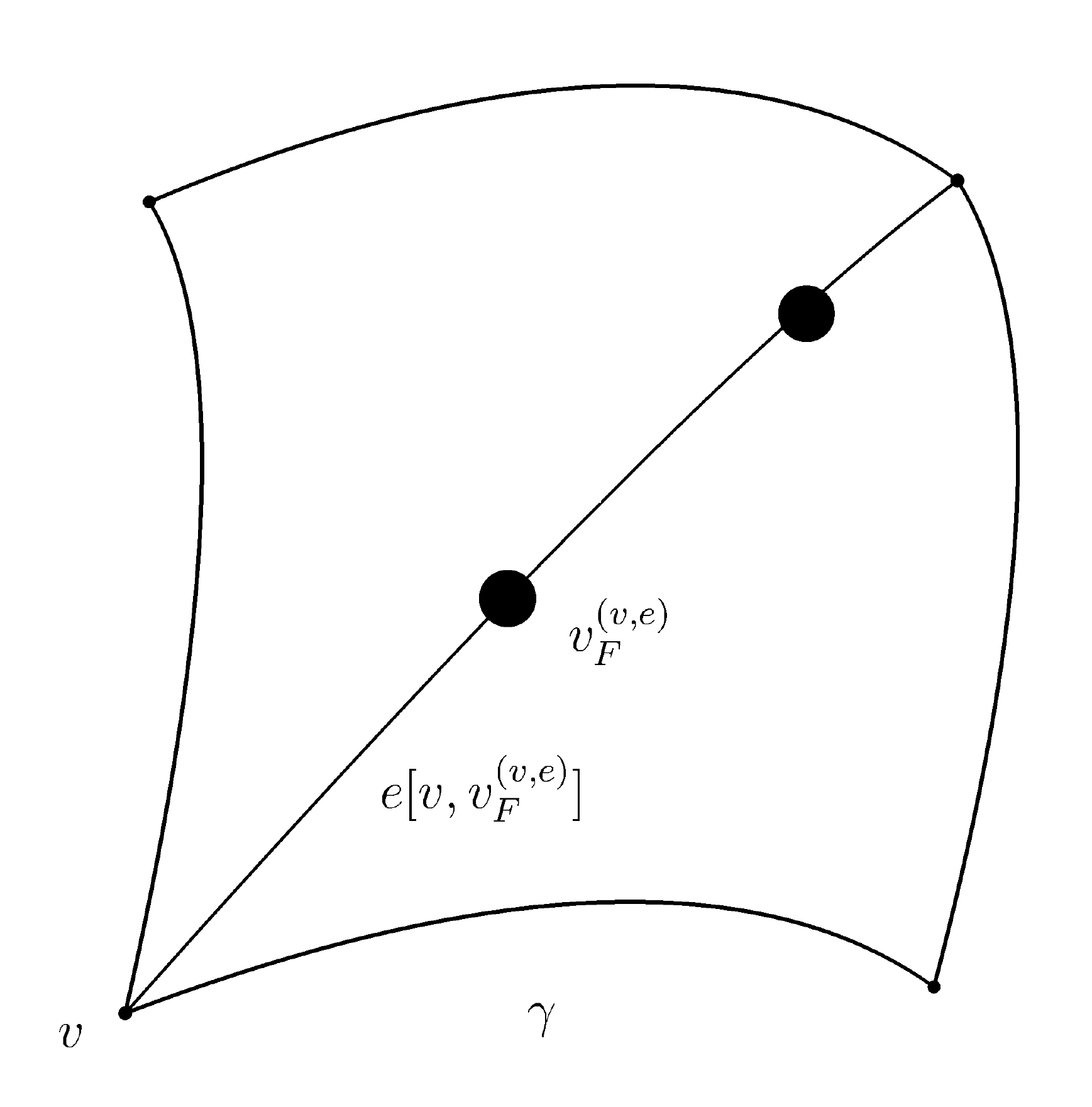}
\caption{An example of graphs $\gamma$ with $v_F^{(v,e)}$ being the closest fermionic vertex to $v$ in $e\in E(\ker(\gamma))$, and $e[v_F^{(v,e)},v]\subset e$ is the edge of $\gamma$ from $v$ to $v_F^{(v,e)}$.}\label{fig:gammappp}
\end{figure}

For the case with $e\subset \phi_o(\gamma)$, $\left\langle\hat{\mathfrak h}_1(v,e;\delta)^\dagger(\phi_o\star\Psi_\gamma)\middle|\Phi_{\gamma'}\right\rangle\neq 0$ requires  that the graph  of  $\hat{\mathfrak h}_1(v,e;\delta)^\dagger(\phi_o\star\Psi_\gamma)$ is equal to $\gamma'$. Since $\delta $ is chosen to be sufficiently small, in the graph $\gamma'$ there is no fermionic vertex on the segment $e(v,\delta)\subset e\subset\gamma'$, and $t_{e(v,\delta)}\in\gamma'$ is also not a fermionic vertex. Therefore, for  $\left\langle\hat{\mathfrak h}_1(v,e;\delta)^\dagger(\phi_o\star\Psi_\gamma)\middle|\Phi_{\gamma'}\right\rangle\neq 0$, there cannot exist any fermionic vertices in $e(v,\delta)\cup\{t_{e(v,\delta)}\}$, where $e(v,\delta)$ and $t_{e(v,\delta)}$ now are thought of as a segment and a vertex in the graph of $\hat{\mathfrak h}_1(v,e;\delta)^\dagger(\phi_o\star\Psi_\gamma)$. This conclusion has two meanings. At first, $\phi_o\in \diff'$ must be the diffeomorphism which moves the closest fermionic vertex to $v$  in $e\in \gamma$  to $t_{e(v,\delta)}$ so that $\hat\theta(t_{e(v,\delta)})$ can kill it. The closest fermionic vertex to $v$ in $e\in \gamma$ will be denoted by $v_F^{(v,e)}$, and  Fig. \ref{fig:gammappp} gives an illustration of $v_F^{(v,e)}$. Second, it means that the fermion state at $v_F^{(v,e)}$ cannot be $|1,1\rangle_{v_F^{(v,e)}}$ because it cannot be annihilated completely by $\hat\theta(t_{v,\delta})$. 
For the second point, we introduce a projection $\hat{\mathbb P}_{v_F}$ on the fermion Hilbert space $\mathcal H_{v_F}$ such that
 \begin{equation}
 \hat{\mathbb P}_{v_F}|i_1,i_2\rangle=\left\{
\begin{array}{cc}
0,&i_1=1=i_2,\\
|i_1,i_2\rangle,&\text{ otherwise. }
\end{array}
 \right.
 \end{equation}
 The first point makes us decompose $\phi_o$ as follows. Let $\gamma_\star$ be the graph of $\gamma'$ with $v_F^{(v,e)}$ dropped, and $v\in \gamma$ promoted to a new fermionic vertex if it is not. Then, we find a diffeomorphism  $\phi_1\in\diff_{V(\ker(\gamma))}/\diff_{\gamma_\star}=\diff_{V(\ker(\gamma_\star))}/\diff_{\gamma_\star}$ such that $\phi_1(\gamma_\star)=\gamma'$. With $\phi_1$, we find $\phi_2\in \diff_{\phi_1(\gamma_\star)}/\diff_{\gamma}$ which moves $v_F^{(v,e)}$ to $t_{e(v,\delta)}$. One thus gets $\phi_o=\phi_2\circ\phi_1$. Introducing $\widehat{\mathfrak H}_v(e)$ as
 \begin{equation}\label{eq:keyHve}
 \widehat{\mathfrak H}_v(e):=\left\{
\begin{array}{cl}
\hat J_i^{v,e} \hat\theta^\dagger(v)\sigma^ih_{e[v_F^{(v,e)}, v]}^{-1}\hat\theta(v_F^{(v,e)})\hat{\mathbb P}_{v_F^{(v,e)}}
,& \text{ if } v_F^{(v,e)}  \text{ exists}.\\
0&\text{\ otherwise}
\end{array}
\right.
\end{equation}
where $e[v_F^{(v,e)}, v]$ denotes the edge from $v$ to $v_F^{(v,e)}$. We thus get
 \begin{equation}\label{eq:final0}
 \left\langle\hat{\mathfrak h}_1(v,e;\delta)^\dagger(\phi_o\star\Psi_\gamma)\middle|\Phi_{\gamma'}\right\rangle=\left\langle \phi_1\star(\widehat{\mathfrak H}_v(e)\Psi_\gamma)\middle|\Phi_{\gamma'}\right\rangle,
 \end{equation}
 where the right hand side is independent of $\delta$, and $\phi_2(\gamma')=\gamma'$   is used. Using \eqref{eq:final0}, we finally get
 \begin{equation}\label{eq:final1}
 \lim_{\delta\to 0}I^{(1)}_{v}(\gamma',\delta)=\sum_{\phi\in\diff_{V(\gamma)}/\diff_{\gamma_\star}}\left\langle\phi\star \widehat{\mathfrak H}_v\Psi_\gamma\middle|\Phi_{\gamma'}\right\rangle
  \end{equation}
  with
  \begin{equation}\label{eq:Hvpsi}
  \widehat{\mathfrak H}_v\Psi_{\gamma}=\sum_{e\text{ at }v}\widehat{\mathfrak H}_v(e)\Psi_{\gamma}, \forall \Psi_\gamma\in\mathcal S_\gamma^{\rm irr}. 
  \end{equation}
 Eq. \eqref{eq:final1} leads to
 \begin{equation}
 \lim_{\delta\to 0}\left(\hat{\mathfrak h}_1(v;\delta)^*\eta(\Psi_\gamma)\right|=\left(\eta(\hat{\mathfrak H}_v\Psi_\gamma)\right|
 \end{equation}
Finally, taking advantage of $\lim_{\delta\to 0}\hat{\mathfrak{h}}_2^*\eta(\Psi_\gamma)=\eta(\hat{\mathfrak{h}}_2   \Psi_\gamma)$, we have
  \begin{equation}\label{eq:H1daggerstatrresult}
\lim_{\delta\to 0}(\widehat{H^{(1)}(v;\delta)}{})^*\eta(\Psi_\gamma)=\kappa\hbar\beta N(v) \eta\left[\left((\hat{\mathfrak H}_v-\sum_{e\text{ at } v}\hat J_i^{v,e}\hat\theta^\dagger(v)\sigma^i\hat\theta(v)\right)\Psi_\gamma  \right].
\end{equation}

Now let us consider the operator $\widehat{H^{(2)}(v;\delta)}$. An issue on defining $\lim\limits_{\delta\to 0} \widehat{H^{(2)}(v;\delta)}$ is the operator $\hat H_{E,v}$ comprising $ \widehat{H^{(2)}(v;\delta)}$. In this paper, we will employ the work \cite{alesci2015hamiltonian} to define $\hat H_{E,v}$ in $\mathcal H_{\rm vtx}$. More precisely, $\hat H_{E,v}$ is given  by $\hat H_{E,v}=  {\widehat{\sqrt{V_v^{-1}}}(\hat F_{v}+\hat F_{v}^\dagger)\widehat{\sqrt{V_v^{-1}}}}$ with 
\begin{equation}
\hat  F_v=-2\sum_{e,e' \text{ at } v}\epsilon(\dot e,\dot e')\epsilon^{ijk} \tr(h_{\alpha_{ee'}}\tau_k)\hat J_i^{v,e}\hat J_k^{v,e'}
\end{equation}
where $\alpha_{ee'}$ is a loop tangent to the two edges $e$ and $e'$ at the vertex $v$ up to orders $k_{e}+1$ and $k_{e'}+1$ respectively with $k_e$ and $k_{e'}$ being respectively the orders of tangentiality of $e$ and $e'$ at $v$ (see \cite{alesci2015hamiltonian} 
  for more details.) By this definition, $\hat  F_v$ will change the graph by adding a loop at $v$ and, thus, $\hat F_v^\dagger$, by removing a loop. 

 A subtlety here is that the vertex Hilbert space defined in \cite{alesci2015hamiltonian}  is a little different from ours. In \cite{alesci2015hamiltonian}, the authors defines the vertex Hilbert space with diffeomorphisms  preserving $V(\gamma)$ while our work  considers the diffeomorphisms preserving $V(\ker(\gamma))$. Regardless of this difference, the operator $\hat H_{E,v}$ introduced in \cite{alesci2015hamiltonian} , for $v\in V(\ker(\gamma))$, is well-defined in our vertex Hilbert space $\mathcal  H_{\rm vtx}$. Indeed, due to the operator $  {\widehat{\sqrt{V_v^{-1}}}}$ in  $\widehat{\tilde H^{(2)}(v;\delta)}$, we do not need to  consider $\hat H_{E,v}$ for removable $v$, i.e. $v\notin V(\ker(\gamma))$. According  to this discussion, for $v\in V(\ker(\gamma))$, we have
\begin{equation}\label{eq:H2starresult}
\begin{aligned}
\lim_{\delta\to 0}\widehat{H^{(2)}(v;\delta)}\eta(\Psi_\gamma)=\frac{\kappa_1(v)}{2i\kappa\hbar\beta^2}N(v)\eta\left[\left(\hat  H_{E,v}\hat V_v-\hat V_v\hat H_{E,v}\right)\hat\theta^\dagger(v)\hat\theta(v)\Psi_\gamma\right],
\end{aligned}
\end{equation}
where $\kappa_1(v)$ is introduced  in \cite{alesci2015hamiltonian} to remove the dependence on  the partition.  
  
Finally, for the operator $\widehat{H^{(3)}(v;\delta)}$, since $\widehat{H^{(3)}(v;\delta)}$ for $v\in V(\ker(\gamma))$ is independent of $\delta$ and $\diff_{V(\ker(\gamma))}$ invariant. we have
\begin{equation}\label{eq:H3starresult}
\lim_{\delta\to 0}\left(\widehat{H^{(3)}(v;\delta)}{}^*\eta(\Psi_\gamma)\right|=\left(\eta(\hat{H}^{(3)}_F(v)\Psi_\gamma)\right|
\end{equation}
with 
\begin{equation}
\hat{H}^{(3)}_F(v):=\kappa\hbar\beta N(v)\left(\sum_{e \text{ at } v_C}\hat J_i^{v_C,e}\right)\hat\theta^\dagger(v)\sigma^i\hat\theta(v).
\end{equation}

Let us summarize our results. According to the above discussion, we have
\begin{equation}
\hat{A}_F\eta(\Psi_\gamma):=\lim_{\delta\to 0} \widehat{H(\delta)}{}^*\eta(\Psi_\gamma)=\sum_{v\in V(\gamma)}\eta\left(\widehat{\sqrt{V_v^{-1}}}\hat H_v\widehat{\sqrt{V_v^{-1}}}\Psi_\gamma\right)
\end{equation}
where $\hat H_v$ is given by
\begin{equation}
\hat H_v=i\widehat{H^{(1)}(v)}-\frac{\beta}{2}\widehat{H^{(2)}(v)}-\frac{1+\beta^2}{2\beta}\widehat{H^{(3)}(v)}-\beta\widehat{H^{(1)}(v)},
\end{equation}
with
\begin{equation}\label{eq:h123final}
\begin{aligned}
\widehat{H^{(1)}(v)}&=\kappa_2(v)\Pl^2\beta N(v)\left(\hat{\mathfrak H}_v-\sum_{e\text{ at } v}  \hat\theta^\dagger(v)\sigma^i\hat\theta(v) \hat J_i^{v,e}\right),\\
\widehat{H^{(2)}(v)}&=\frac{\kappa_1(v)}{2i\Pl^2\beta^2}N(v)\left(\hat  H_{E,v}\hat V_v-\hat V_v\hat H_{E,v}\right)\hat\theta^\dagger(v)\hat\theta(v),\\
\widehat{H^{(3)}(v)}&=\kappa_3(v)\Pl^2\beta N(v)\left(\sum_{e \text{ at } v}\hat J_i^{v,e}\right)\hat\theta^\dagger(v)\sigma^i\hat\theta(v),
\end{aligned}
\end{equation}
according to  \eqref{eq:H1daggerstatrresult}, \eqref{eq:H2starresult} and  \eqref{eq:H3starresult} respectively. Here we again introduce the parameters $\kappa_2$ and $\kappa_3$ as in \cite{ashtekar1997quantumII} in order to remove the dependence on  the partition. Finally, because of \eqref{eq:limitH}, we finally define the  fermion Hamiltonian operator $\hat H_F[N]$ on $\mathcal H_{\rm vtx}$ as
\begin{equation}
\hat H_F[N]=\frac{1}{2}(\hat{A}_F+\hat{A}_F^\dagger).
\end{equation}

 Let $\hat{\mathfrak{H}}_v'$ on $\mathcal H_{\rm vtx}$ be the operator  defined by
\begin{equation}
\hat{\mathfrak{H}}_v'\eta(\Psi_\gamma)=\eta(\hat{\mathfrak{H}}_v\Psi_\gamma). 
\end{equation} 
Then,  the only nontrivial  term in $A_F^\dagger$ is the operator $(\hat{\mathfrak{H}}_v')^\dagger$. By the definition, $\hat{\mathfrak H}_v$ is composed of $\hat{\mathfrak  H}_v(e)$ which annihilates the fermionic vertex contained in $e$ and closest to $v$.  Thus, the operator $(\hat{\mathfrak{H}}_v')^\dagger$ contains the  operators, each of which is associated to an edge $e$ at $v$ and promotes a point $v_F\in e$ to a fermionic vertex  such that  $v_F$ becomes the closest to $v$ among the fermionic vertices contained in $e$.  To  be  more precise, let us define an operator $\hat{\mathfrak H}_v^\#(e)$ (refer to \eqref{eq:keyHve}) as
\begin{equation}\label{eq:corresponding H1}
\widehat{\mathfrak H}_v^\#(e)=\chi_{+}(|\hat J^{v,e}|^2)\hat \theta^\dagger(\mathfrak{v}_{e,v}) h_{e(\mathfrak{v}_{e,v})}\sigma^i\hat\theta(v) \hat J_i^{v,e}
\end{equation}
where $\chi_{+}$ denotes the characteristic function of $[\frac34,\infty)$, $|\hat J^{v,e}|^2=\delta^{kl}\hat J^{v,e}_k\hat J^{v,e}_l$,  
$\mathfrak{v}_{e,v}\in e$ is arbitrary point between $v$ and the closest fermionic vertex to $v$ in $e$, and $e(\mathfrak{v}_{e,v})\subset e$ is the segment starting from $v$ and  ending at  $\mathfrak{v}_{e,v}$. Here the factor $\chi_{+}(|\hat J^{v,e}|^2)$ ensures that $\widehat{\mathfrak H}_v^\#(e)$ cannot erase the segment $e(\mathfrak{v}_{e,v})\subset e$. In other words, $\widehat{\mathfrak H}_v^\#(e)$, acting on a state with spin $\frac12$ on $e(\mathfrak{v}_{e,v})$, will  only change the spin $\frac12$ to $1$. Moreover, $\mathfrak{v}_{e,v}$ is defined such that it becomes closest  to $v$ among the fermionic vertices in $e$ after the action  of $\widehat{\mathfrak H}_v^\#(e)$.  With $\widehat{\mathfrak H}_v^\#(e)$, $(\hat{\mathfrak{H}}_v')^\dagger$ is given by
\begin{equation}
(\hat{\mathfrak{H}}_v')^\dagger\eta(\Psi_\gamma)=\eta\left(\sum_{e\text{ at } v}\widehat{\mathfrak H}_v^\#(e)\Psi_\gamma  \right).
\end{equation}
Note that even though $\mathfrak{v}_{e,v}$ is not uniquely determined by its definition, the operator $(\hat{\mathfrak{H}}_v')^\dagger$ on $\mathcal H_{\rm vtx}$ is independent of the choice of  $\mathfrak{v}_{e,v}$, due to the averaging  operation $\eta$. Moreover, $\chi_+(|\hat J^{v,e}|^2)$  is introduced in \eqref{eq:corresponding H1} because \eqref{eq:Hvpsi} implies 
\begin{equation}
\widehat{\mathfrak H}_v=\sum_{e\text{ at } v}\hat{\mathfrak H}_v(e)\chi_+(|\hat J^{v,e}|^2).
\end{equation}
Applying  $\widehat{\mathfrak H}_v^\#(e)$, $\hat A_F^\dagger$ reads
\begin{equation}
\hat{A}_F^\dagger\eta(\Psi_\gamma)=\sum_{v\in V(\gamma)}\eta\left(\widehat{\sqrt{V_v^{-1}}}\hat H_v^\#\widehat{\sqrt{V_v^{-1}}}\Psi_\gamma\right)
\end{equation}
where $\hat H_v^\#$ is 
\begin{equation}
\hat H_v^\#=-i\widehat{H^{(1)}(v)}{}^\#-\frac{\beta}{2}\widehat{H^{(2)}(v)}-\frac{1+\beta^2}{2\beta}\widehat{H^{(3)}(v)}-\beta\widehat{H^{(1)}(v)}{}^\#,
\end{equation}
with $\widehat{H^{(2)}v)}$ and $\widehat{H^{(3)}(v)}$ given in \eqref{eq:h123final}, and $\widehat{H^{(1)}(v)}{}^\#$ given by
\begin{equation}
\begin{aligned}
\widehat{H^{(1)}(v)}{}^\#&=\kappa_2(v)\Pl^2\beta N(v)\left(\sum_{e\text{ at } v}\widehat{\mathfrak H}_v^\#(e)-\sum_{e\text{ at } v}  \hat\theta^\dagger(v)\sigma^i\hat\theta(v) \hat J_i^{v,e}\right).
\end{aligned}
\end{equation}

Let us complete this section with a discussion on the intuitive picture led by the action of $\widehat{\mathfrak H}_v(e)$ and $\widehat{\mathfrak H}_v^\#(e)$. By definition, $\widehat{\mathfrak H}_v^\#(e)$ creates a fermionic vertex $v_F\in e$ so that $v_F$ will be closest to $v$ and carries states $a|1,0\rangle+b|0,1\rangle$.  Simultaneously, $\widehat{\mathfrak H}_v^\#(e)$  changes the fermion state $|i_1,i_2\rangle_v$ at $v$ in such a way that $|1,1\rangle_v$ becomes $ c|1,0\rangle_v+d|1,0\rangle_v$ for some constant $c$ and $d$, and $\alpha |1,0\rangle_v+\beta|0,1\rangle_v$ for arbitrary $\alpha$ and $\beta$ is changed to $ |0,0\rangle_v$. Moreover, because of the holonomy operator $h_{e(\mathfrak{v}_{e,v})}$ and flux operator $\hat J_i^{v,e}$ in $\widehat{\mathfrak H}_v^\#(e)$, the spin on the segment $e(\mathfrak{v}_{e,v})\subset  e$ and the intertwiner at $v$ are changed.  These results can be summarized as that the operator $\widehat{\mathfrak H}_v^\#(e)$ moves a fermion at $v$ to $v_F$, and changes the geometry around $v$ simultaneously. For the operator $\widehat{\mathfrak H}_v(e)$, it reverses this procedure.  $\widehat{\mathfrak H}_v(e)$ moves a fermion at the fermionic vertex $v_F^{(v,e)}$, i.e., the closest fermionic vertex to $v$, to the vertex $v$, and changes the geometry around $v$ simultaneously. Moreover, because of the projection operator $\hat{\mathbb P}_{v_F}$ in $\widehat{\mathfrak H}_v(e)$ (see  \eqref{eq:keyHve}), when the fermionic vertex $v_F^{(v,e)}$ carries a fermion state $|1,1\rangle_{v_F^{(v,e)}}$, the fermion at $v_F^{(v,e)}$ cannot be moved by the operator $\widehat{\mathfrak H}_v(e)$.  To see the consequence of this fact, let us imagine an edge $e$ with both the source $s_e$ and the middle point $v_F^{(v,e)}$ being the fermionic vertices, where $v_F^{(v,e)}$ carries a fermion state, says, $|1,0\rangle_{v_F^{(v,e)}}$. Acted by $\widehat{\mathfrak H}_{s_e}^\#(e)$, the fermion at $s_e$ will be moved to some point $v_F'$ between $s_e$ and $v_F^{(v,e)}$. Then, acted by   $\widehat{\mathfrak H}_{t_e}(e)$ twice, the fermion at $v_F'$ will be moved to $t_e$. Now, suppose that  $v_F^{(v,e)}$ carries the state $|1,1\rangle_{v_F^{(v,e)}}$. Then, acted by $\widehat{\mathfrak H}_{s_e}^\#(e)$, the fermion at $s_e$ will be again moved to $v_F'$. However,  the fermion at $v_F'$ cannot be moved to $t_e$ after acted by $\widehat{\mathfrak H}_{t_e}(e)$, due to the operator $\hat{\mathbb P}_{v_F^{(v,e)}}$ in $\widehat{\mathfrak H}_{t_e}(e)$. Intuitively, in the second situation, the fermion at $s_e$ is confined around $s_e$ by the fermion state $|1,1\rangle_{v_F^{(v,e)}}$ at $v_F^{(v,e)}$. 
The above picture tells  how a fermion moves in loop quantum spacetime and influences the background quantum geometry  in  the LQG framework.

\section{summation and outlook}\label{sec:five}
This work is concerned about the  model of fermion field coupled to LQG. The Gauss and Hamiltonian constraints in this model are studied in details. In the solution to the Gauss constraint, fermion spins and the gravitational spin network intertwine with each other so that the fermion spins contribute to the volume of the spin network vertices. Consequently, the closure condition encoded in the Gauss  constraint will no longer be satisfied for the gauge invariant state with non-vanishing  fermion spins. In other words, the faces dual to the edges at a fermionic vertex with non-vanishing fermion spin could not form a closed polyhedron, and the area defect of this unclosed polyhedron is filled by the fermion spin. Consequently, in contract to pure-gravity case, a 3-valence gauge invariant vertex with non-vanishing fermion spin will get non-vanishing volume from the fermion spin. The volume of this type of vertices is computed in details. 

For the Hamiltonian constraint,  the regularization and quantization procedures are presented in details. There are several remarkable issues on the Hamiltonian constraint operator. At first, in order to take the limit of the regularized expression as the regulator approaches 0, we introduce the vertex Hilbert space. By definition, the vertex Hilbert space is the space of cylindrical functions averaged with the diffeomorphisms preserving the unremovable vertices.  Thus, the states in the vertex Hilbert space are partially diffeomorphism invariant. The vertex Hilbert space is the dual space to the space of cylindrical functions. Then, the regularized Hamiltonian operator $\widehat{H(\delta)}$ can be promoted as an operator $\widehat{H(\delta)}{}^*$ therein by duality. 
Due to the diffeomorphism invariant feature of the vertex Hilbert space, the operators $\widehat{H(\delta)}{}^*$ for different values of the regulator $\delta$ are identical. Thus, the limit of $\widehat{H(\delta)}{}^*$ as  $\delta$ approaches 0 can be taken. Moreover,  by introducing the vertex Hilbert space, we fix several other problematic issues. Classically,  the Hamiltonian can be divided into two parts as $H_{\mathcal C_\epsilon}$ plussing its complex conjugate $\overline{H_{\mathcal C_\epsilon}}$. $H(\delta)$ can be promoted to the operator $\widehat{H(\delta)}$ which is cylindrical consistent and diffeomorphism covariant. However, its adjoint $\widehat{H(\delta)}{}^\dagger$, as a candidate of the operator corresponding to $\overline{H_{\mathcal C_\epsilon}}$,  is not densely defined, because  $\widehat{H(\delta)}$ could change graphs by erasing some segments of edges in graphs. These problems on $\widehat{H(\delta)}{}^\dagger$ are finally solved by introducing the vertex Hilbert space to define limit. In the vertex Hilbert space,  the operator $\hat A_F^\dagger$, as the limit of $H(\delta)$, performs in a way such that 
 a projection is left multiplied in $\hat A_F^\dagger$ as a factor. Then, if $\hat A_F^\dagger$ erases segments of edges, the projection will annihilate the resulting state, so that the adjoint of $\hat A_F^\dagger$ is densely defined.  
 Finally, in the Hamiltonian constraint operator, there are  the operators $\widehat{\mathfrak H}_v^\#(e)$ and $\widehat{\mathfrak H}_v(e)$ involved. These two operators tell  how a fermion moves in LQG  spacetime and influences the background loop quantum geometry. According  to our results, the operator $\widehat{\mathfrak H}_v^\#(e)$ moves a fermion at the vertex $v$ to a point $v_F\in e$ so that $v_F$ becomes the closest fermionic vertex to $v$, and simultaneously changes the spin on the segment connecting $v$ to $v_F$ of $e$ as well as the intertwiner at $v$. This procedure will be reversed by the operator $\widehat{\mathfrak H}_v(e)$, which moves a fermion,  located at the fermionic vertex $v_F^{(v,e)}$ closest to $v$ in $e$, to the vertex $v$,  and simultaneously changes the spin on the segment connecting $v$ to $v_F^{(v,e)}$ in $e$ as well as the intertwiner at $v$. In addition,  $\widehat{\mathfrak H}_v(e)$ is defined to contain a projection operator $\hat{\mathbb P}_{v_F^{(e,v)}}$. As a consequence of this operator, the fermion located at, says, $s_e$ will be confined around $s_e$ by the state $|1,1\rangle_{v_F^{(v,e)}}$ located at $v_F^{(v,e)}\in e$.

Even though the current work is concerned about the graph changing feature, the framework can be easily adapted to define a graph preserving version of the Hamiltonian constraint operator. Then one can apply this graph preserving operator to the lattice model of fermion coupled to LQG, so that some open issues in lattice fermion field theory can be employed and studied. Moreover, the properties of the fermion Hamiltonian operator are still not well understood although we have discussed some of them. All of these will be left as our future works. 

\section*{Acknowledgements}

This work is benefited greatly from the numerous discussions with Muxin Han, and supported by the Polish Narodowe Centrum Nauki, Grant No. 2018/30/Q/ST2/00811, and NSFC with Grants No. 11961131013 and No. 11875006.

\appendix
\section{Hamiltonian analysis for fermion field}\label{app:hamiltoniananalysisFermion}
Define $P_\pm=\frac{1\pm \gamma^5}{2},$ one has
\begin{equation}
\begin{aligned}
P_\pm^2=P_\pm,\ P_+P_-=P_-P_+=0,\ P_\pm\gamma^\mu=\gamma^\mu P_\mp.
\end{aligned}
\end{equation}
Therefore, we have
\begin{equation}
\begin{aligned}
\overline\Psi\gamma^I e_I^\mu P_{\pm} \nabla_\mu\Psi=(\Psi_\pm)^\dagger\gamma^0\gamma^I e_I^\mu \Psi_\pm+\frac{1}{4}(\Psi_\pm)^\dagger\gamma^0\gamma^I e_I^\mu \omega_{\mu KL}\gamma^K\gamma^L\Psi_\pm
\end{aligned}
\end{equation}
with $\Psi_{\pm}:=P_\pm\Psi$. Let us choose the Weyl basis of the $\gamma$ matrices 
\begin{equation}
\gamma^0 = \begin{pmatrix} 0 &  i \mathbbm 1_2 \\ i\mathbbm 1_2 & 0 \end{pmatrix},\quad \gamma^k = \begin{pmatrix} 0 & i\sigma^k \\ -i\sigma^k & 0 \end{pmatrix},\quad \gamma^5 = \begin{pmatrix} -I_2 & 0 \\ 0 & I_2 \end{pmatrix}.
\end{equation}
Then $\Psi_{\pm }$ take the form $\Psi_{-}=(\psi,0)^T$ and  $\Psi_{+}=(0,\eta)^T$. We thus get
\begin{equation}\label{eq:gammatosigma1}
\begin{aligned}
&\overline\Psi\gamma^I e_I^\mu \nabla_\mu\Psi=-\psi^\dagger\overline\sigma^I e_I^\mu\partial_\mu \psi+\frac{1}{4}\psi^\dagger\overline\sigma^I e_I^\mu \omega_{\mu KL}\sigma^K\overline\sigma^L\psi-\eta^\dagger\sigma^I e_I^\mu \eta+\frac{1}{4}\eta^\dagger\sigma e_I^\mu \omega_{\mu KL}\overline\sigma^K\sigma^L\eta
\end{aligned}
\end{equation}
with $\sigma^I=(\mathbbm 1,\text{Pauli matrax}^i)$ and $\overline\sigma^I=(\mathbbm 1,-\text{Pauli matrax}^i)$. 

Performing the 3+1-decomposition $\mathcal M=\mathbb R\times\Sigma$, one has 
\begin{equation}\label{eq:3plus1e}
e_I^\mu=e^\nu_I q_\nu^\mu-n^\mu n_I
\end{equation}
where $q_\nu^\mu$ is the projection to $\Sigma$ and $n^\mu=(t^\mu-N^\mu)/N$ with $N$ and $N^\mu$ being the lapse function and the shift vector respectively, and $t^\mu$ being some time evolution vector field given by $t^\mu \partial_\mu t=1$.  Substituting \eqref{eq:3plus1e} into \eqref{eq:gammatosigma1}, we have 
\begin{equation}
\begin{aligned}
\overline\Psi e_I^\mu \gamma^I\nabla_\mu\Psi=\left(\psi^\dagger e_i^a\sigma^i\mathcal D^+_a\psi-\eta^\dagger e_i^a\mathcal D^-_a\eta \right)-\frac{1}{N}(t^\mu-N^\mu)\left(\psi^\dagger \mathcal D_\mu^+ \psi+\eta^\dagger  \mathcal D_\mu^-\eta\right),
\end{aligned}
\end{equation}
where we defined
\begin{equation}\label{eq:calD}
\mathcal D^\pm_a=\partial_a+(\Gamma_a^m\mp K_a^m)\tau_m=:\partial_a+\mathcal A^\pm_a.
\end{equation}
Defining 
\begin{equation}
\nabla_a=\partial_a+\Gamma_a^m\tau_m
\end{equation}
we can express the action of the fermion field in terms of $K_a^i$ and $\nabla_a$ explicitly, which reads
\begin{equation}\label{eq:actionSF}
\begin{aligned}
S_F=&-\frac{i}{2}\int_{\mathcal M}\dd^4 x (\overline\Psi\gamma^I e_I^\mu \nabla_\mu\Psi-c.c)\\
=&\frac{i}{2}\int\dd^4 x \sqrt{q}\Bigg( \left(\psi^\dagger \partial_t\psi+\eta^\dagger \partial_t\eta-c.c\right)+2 \Gamma_{t m}\left(\psi^\dagger \tau^m \psi+\eta^\dagger\tau^m\eta\right)-N^a\left(\psi^\dagger \nabla_a\psi+\eta^\dagger \nabla_a\eta-(\nabla_a\psi)^\dagger\psi-(\nabla_a\eta)^\dagger\eta \right)\\
&-\frac{N}{\sqrt{q}}\Big[\psi^\dagger E_i^a\sigma^i \nabla_a\psi-(\nabla_a\psi)^\dagger E_i^a\sigma^i\psi  +2 \psi^\dagger [E^a, K_a]\psi-\eta^\dagger E_i^a\sigma^i\nabla_a\eta+(\nabla_a\eta)^\dagger E_i^a\sigma^i\eta+2\eta^\dagger [E^a, K_a]\eta\Big]\Bigg)
\end{aligned}
\end{equation}
Define $\xi_A=\sqrt[4]{q}\psi_A$ and $\nu_A=\sqrt[4]{q}\eta_A$ with $A=\pm \frac12$. Eq. \eqref{eq:actionSF} implies the following non-vanishing anti-Poisson bracket, 
\begin{equation}
\begin{aligned}
\{\xi_A(x),\xi_B^\dagger(y)\}_+=-i \delta_{AB}\delta(x,y)\\
\{\nu_A(x),\nu_B^\dagger(y)\}_+=-i \delta_{AB}\delta(x,y)
\end{aligned}
\end{equation}

For the gravitational parts, the action is
\begin{equation}\label{eq:actionSH}
\begin{aligned}
S_H=\frac{1}{\kappa}\int\dd^4 x\,(E^a_i \mathcal L_tK_a^{i}+\frac{1}{2\sqrt{q}}N E^a_i E^b_j\Omega_{ab}^{ij}+(t\cdot\Gamma)_m\epsilon^{kl m} K_{ak}E^a_l+2N^b E^a_i \nabla_{[a}K_{b]}^i)
\end{aligned}
\end{equation}
Substituting the expression \eqref{eq:actionSF} and \eqref{eq:actionSH} into the total action $S=S_H+S_F$, one can obtain the constraints governing the classical dynamics which are expressed in terms of $\nabla_a=\partial_a+\Gamma_a^m\tau_m$.  Then taking advantage of $A_a^i=\Gamma_a^i+\beta K_a^i$, one can simplify these constraints in terms of the derivative $D_a=\partial_a+A_a^i\tau_i$. The results are listed as follows.
The total action reads
\begin{equation}
S=S_G+S_F=\int\dd^4 x(\text{simplytic structure terms }-\lambda ^m G_m-N^a H_a -NH).
\end{equation}
The Gaussian constraint is
\begin{equation}
\begin{aligned}
G_m=\frac{1}{\kappa\beta}D_a E^a_l+\frac{1}{2} \sqrt{q}(\psi^\dagger\sigma_m\psi+\eta^\dagger\sigma_m\eta)
\end{aligned}
\end{equation}
The vector constraints is
\begin{equation}
\begin{aligned}
H_a=\frac{1}{\kappa\beta} E^b_i F^i_{ab}+ \frac{i}{2}\sqrt{q}\Big\{\psi^\dagger D_a\psi-(D_a\psi)^\dagger\psi+\eta^\dagger D_a\eta-(D_a\eta)^\dagger\eta \Big\}+\beta K_a^m G_m.
\end{aligned}
\end{equation}
The scalar constraint is
\begin{equation}
\begin{aligned}
H=&H_G+ \frac12 \Big[i(\psi^\dagger E_i^a\sigma^i D_a\psi-(D_a\psi)^\dagger E_i^a\sigma^i\psi)-\beta E^a_i K_a^i \psi^\dagger\psi-\frac{1}{\beta}(1+\beta^2)D_a E^a_i \psi^\dagger\sigma^i\psi-\beta\frac{1}{\sqrt{q}} E^a_iD_a\Big(\sqrt{q} \psi^\dagger\sigma^i\psi \Big)\\
&-i(\eta^\dagger E_i^a\sigma^iD_a\eta-(\nabla_a\eta)^\dagger E_i^a\sigma^i\eta)+\beta E^a_i K_a^i  \eta^\dagger \eta-\frac{1}{\beta}(1+\beta^2)D_a E^a_i\eta^\dagger \sigma^i\eta-\beta\frac{1}{\sqrt{q}} E^a_iD_a\Big(\sqrt{q}\eta^\dagger\sigma^i\eta\Big)\Big]\\
\end{aligned}
\end{equation}
where $H_G$ denote the scalar constraint of pure gravity
\begin{equation}
H_G=\frac{1}{2\kappa\sqrt{q}} E^a_i E^b_j\left(F_{ab}^m\epsilon_m{}^{ij}-2(1+\beta^2)K_{[a}^iK_{b]}^j\right).
\end{equation}

\section{graded vector space and graded algebra}\label{app:graded}
We follow the notions given in \cite{kostant1975graded}.
A vector space $V$ over $\mathbb R$ or $\mathbb C$  is graded (over $\mathbb Z_2$) if there are fixed subspaces $V_0$ and $V_1$ such that $V=V_0\oplus V_1$.
An element $v\in V$ is homogeneous if $v$ is either in $V_0$ or in $V_i$. For all $v\in V_i$ with $v\neq 0$, we define their degree as
\begin{equation}
\grade(v)=i.
\end{equation}
Given two graded vector space $V$ and $W$, the space ${\rm Hom}(V,W)$ of homomorphism from $V$ to $W$ is graded. An element $\alpha\in {\rm Hom}(V,W)$ is said to be homogeneous and of  $\grade(\alpha)$ provided
\begin{equation}
\alpha[V_i]\subset W_{i+\grade(\alpha)\,{\rm mod}\, 2}.
\end{equation}
with $\alpha[V_i]$ denotes the image of $\alpha$ acting on $V_i$. 

An algebra $(A,\cdot)$ is a graded algebra if $A$ is a graded vector space and $A_i\cdot A_j\subset A_{i+j\, \mathrm{mod}\, 2}$ where $A_i\cdot A_j$ denotes the space of elements $a_i\cdot a_j$ for all $a_i\in A_i$ and $a_j\in A_j$.  A graded algebra $A$ is a graded commutative algebra if the product satisfies 
\begin{equation}
x\cdot y=(-1)^{\grade(x)\grade(y)} y\cdot x
\end{equation}
where $x,y\in A$ are homogeneous.  Any commutative algebra $\mathbb A$ is a graded commutative algebra with the grade $\mathbb A_1=\mathbb A$ and $\mathbb A_0=\{0\}$.
An example of the graded commutative algebra is the exterior algebra of some finite vector space $V$, i.e.
\begin{equation}
A=\mathbb R\oplus V\oplus (V\wedge V)\oplus (V\wedge V\wedge V)\oplus\cdots \oplus \bigwedge^n V.
\end{equation}
$A$ is graded as 
\begin{equation}
A_0=\bigoplus_{k=0}\bigwedge^{2k} V,\ A_0=\bigoplus_{k=0}\bigwedge^{2k+1} V.
\end{equation}

A graded algebra $(\mathfrak a,[\cdot,\cdot])$ is a graded Lie algebra if the Lie bracket satisfies 
\begin{itemize}
\item[(1)] $[x,y]=(-1)^{1+\grade(x)\grade(y)}[y,x]$;
\item[(2)] $(-1)^{\grade(x)\grade(z)}[[x,y],z]+(-1)^{\grade(y)\grade(x)}[[y,z],x]+(-1)^{\grade(z)\grade(y)}[[z,x],y]=0$.
\end{itemize}
An operation $\partial$ on a graded algebra $A$ is called a derivative if it satisfies
\begin{equation}\label{eq:leibniz}
\partial(x y)=(\partial x) y+(-1)^{\grade(\partial)\grade(x)}x(\partial y)
\end{equation}
where $\grade(\partial)$ is defined by thinking of it as a homomorphism on $A$. It can be checked that the operator $[x,\cdot]$ on a graded Lie algebra $A$ for all $x\in A$ is a derivative.



\end{document}